\documentclass[apj,iop]{emulateapj}

\usepackage{natbib}
\usepackage{longtable}
\usepackage{hyperref}

\shorttitle{Structures of Dwarf Galaxies with AGNs}
\shortauthors{Kimbrell et al.}

\begin{document}

\title{The Diverse Morphologies and Structures of Dwarf Galaxies Hosting Optically-Selected Active Massive Black Holes}

\author{Seth J. Kimbrell}
\affil{eXtreme Gravity Institute, Department of Physics, Montana State University, MT 59715, USA}
\email{seth.kimbrell@montana.edu}

\author{Amy E. Reines}
\affil{eXtreme Gravity Institute, Department of Physics, Montana State University, MT 59715, USA}

\author{Zachary Schutte}
\affil{eXtreme Gravity Institute, Department of Physics, Montana State University, MT 59715, USA}

\author{Jenny E. Greene}
\affil{Department of Astrophysical Sciences, Princeton University, Princeton, NJ 08544, USA}

\and

\author{Marla Geha}
\affil{Department of Astronomy, Yale University, New Haven, CT 06520, USA}

\begin{abstract}
We present a study of 41 dwarf galaxies hosting active massive black holes (BHs) using {\it Hubble Space Telescope} observations. The host galaxies have stellar masses in the range of $M_\star \sim 10^{8.5}-10^{9.5}~M_\odot$ and were selected to host active galactic nuclei (AGNs) based on narrow emission line ratios derived from Sloan Digital Sky Survey spectroscopy. We find a wide range of morphologies in our sample including both regular and irregular dwarf galaxies. We fit the {\it HST} images of the regular galaxies using GALFIT and find that the majority are disk-dominated with small pseudobulges, although we do find a handful of bulge-like/elliptical dwarf galaxies. We also find an unresolved source of light in all of the regular galaxies, which may indicate the presence of a nuclear star cluster and/or the detection of AGN continuum. Three of the galaxies in our sample appear to be Magellanic-type dwarf irregulars and two galaxies exhibit clear signatures of interactions/mergers. This work demonstrates the diverse nature of dwarf galaxies hosting optically-selected AGNs. It also has implications for constraining the origin of the first BH seeds using the local BH occupation fraction at low masses -- we must account for the various types of dwarf galaxies that may host BHs.

\end{abstract}

\section{Introduction}
It is well established that massive galaxies host supermassive black holes (BHs) with masses of $M_{\rm BH} \sim 10^{6}-10^{10} M_\odot$ at their centers (\citealt{kormendy,kormendy1995}). Our own Milky Way hosts Sagittarius A*, a BH with a mass of $4 \times 10^6 M_\odot$ \citep{ghez2008}. Much work has gone into studying structural properties, scaling relations, and the possible coevolution of massive galaxies and the BHs they host (see e.g., the review by \citealt{kormendy}). The presence and properties of dwarf galaxies ($M_\star \lesssim 10^{9.5} M_\odot$) hosting massive BHs is not nearly as well studied (see \citealt{reinescomastri2016,greenetal2019} for reviews), with the first systematic search for these objects performed by \citet{reines}.

Lower mass BHs in dwarf galaxies provide a chance to put constraints on BH ``seed" masses and probe their possible formation channels (see e.g., \citealt {volonteri} and \citealt{Greene}). \citet{Mortlock} report observations of a luminous ($6.3 \times 10^{13} L_\odot$) quasar, hosting a $\sim 2 \times 10^9 M_\odot$ BH at a redshift of z=7.085 (corresponding to 0.77 Gyr after the Big Bang). This suggests that the first BH seeds were born in the very early Universe and at least some grew to enormous masses extremely fast. Since we cannot directly observe the small BH seeds at high redshift with current telescopes (\citealt{volonterireines2016,vito}), dwarf galaxies in the low-redshift Universe are our best chance to study BHs that have not grown much compared to the BHs in today's massive galaxies (\citealt{habouzit2016,anglesalcazar}).

\begin{figure*}[!t]
$\begin{array}{cc}
\hspace{.6cm}
{\includegraphics[scale=0.7]{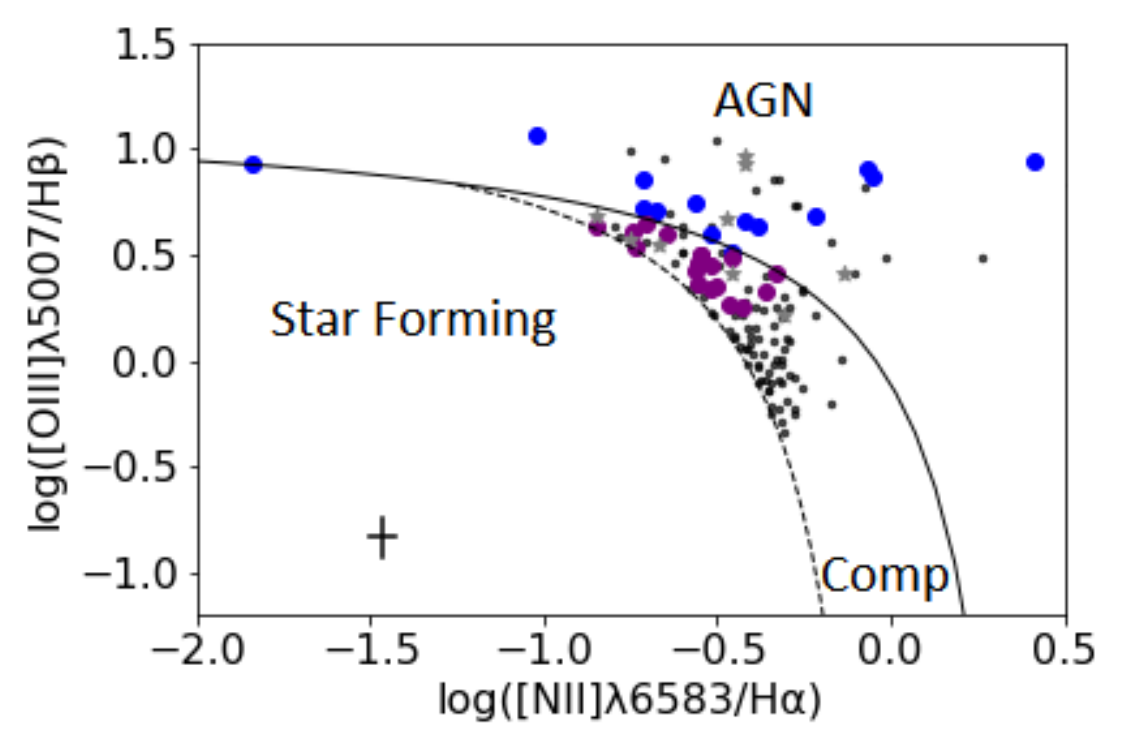}} &
{\includegraphics[scale=0.7]{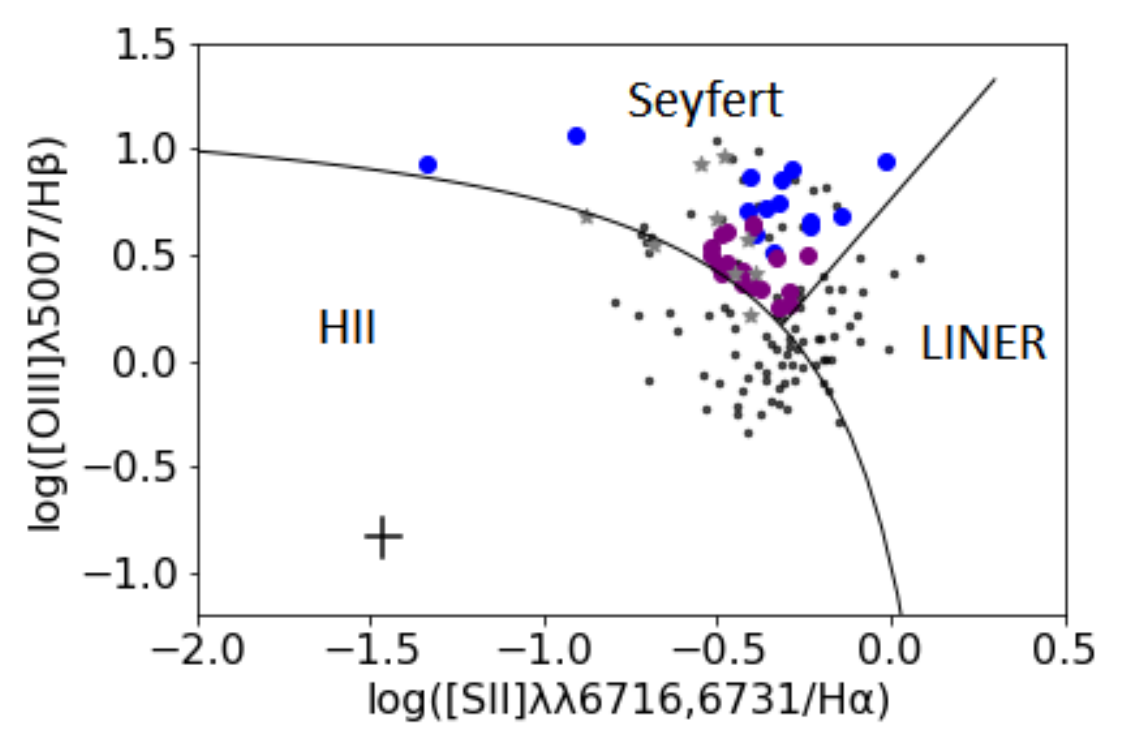}}
\end{array}$
\caption{Narrow emission line diagnostic diagrams showing the location of our sample of dwarf galaxies, which exhibit optical signatures of active massive BHs \citep{reines}. Blue and purple points indicate AGNs and Composites, respectively, with new {\it HST} observations presented here. Grey stars indicate additional galaxies we include in this work from the \citet{reines} sample with {\it HST} imaging previously presented in \citet{schutte}, \citet{baldassare} and \citet{jiang}. Black dots indicate the remaining galaxies in the \citet{reines} sample. Left: [O${\rm III}$]/H$\beta$ vs.\ [N${\rm II}$]/H$\alpha$ diagnostic diagram. The solid line shows the ``maximum starburst" line from stellar photoionization models \citep{KewleyStarburst}. The dashed line is an empirical separation between galaxies that show some contribution from AGN and galaxies dominated by star formation \citep{Kauffmann}. Composite galaxies fall between the dashed and solid lines, and likely indicate contributions from both an AGN and star formation. Right: [O${\rm III}$]/H$\beta$ vs.\ [S${\rm II}$]/H$\alpha$ diagnostic diagram, using the classifications from \citet{kewleyactive}. Typical errors are shown in the lower left corners.}
\label{fig:BPT}
\end{figure*}

In most cases, BHs in dwarf galaxies must be detected as active galactic nuclei (AGN) through radiative signatures, rather than through stellar or gas dynamics (although see \citealt{Nguyen}) since the gravitational sphere of influence of a low-mass BH in a dwarf galaxy is too small to be resolved at distances greater than 4-5 Mpc with current facilities. The first systematic search for AGNs in dwarf galaxies was performed by \citet{reines} using optical spectroscopy from the Sloan Digital Sky Survey (SDSS). They identified 136 dwarf galaxies with narrow emission line ratios indicating the presence of an AGN, some of which also had broad H$\alpha$ emission that was used to estimate BH masses by employing standard viral techniques \citep[e.g.,][]{greenehovirial}. However, as dwarf galaxies have relatively small sizes, very little is known about the detailed morphologies and structures of the host galaxies from ground-based imaging.

In this work, we present {\it Hubble Space Telescope (HST)} imaging of a subset of the \citet{reines} sample. We aim to characterize the structural components of dwarf galaxies hosting active massive BHs and better understand what galactic properties (if any) contribute to or influence the presence of AGNs in dwarf galaxies. We describe our sample of dwarf galaxies and {\it HST} observations in Sections 2 and 3, respectively. Our structural analysis of the galaxy images is presented in Section 4 and our results are given in Section 5. We present a discussion in Section 6, and end with concluding remarks in Section 7. We adopt a Hubble constant of $H_0=73$ km s$^{-1}$ Mpc$^{-1}$ throughout this work, and we report magnitudes in the ST system.

\begin{figure*}[!t]
\begin{center}
$\begin{array}{cc}
{\includegraphics[width=0.42\textwidth]{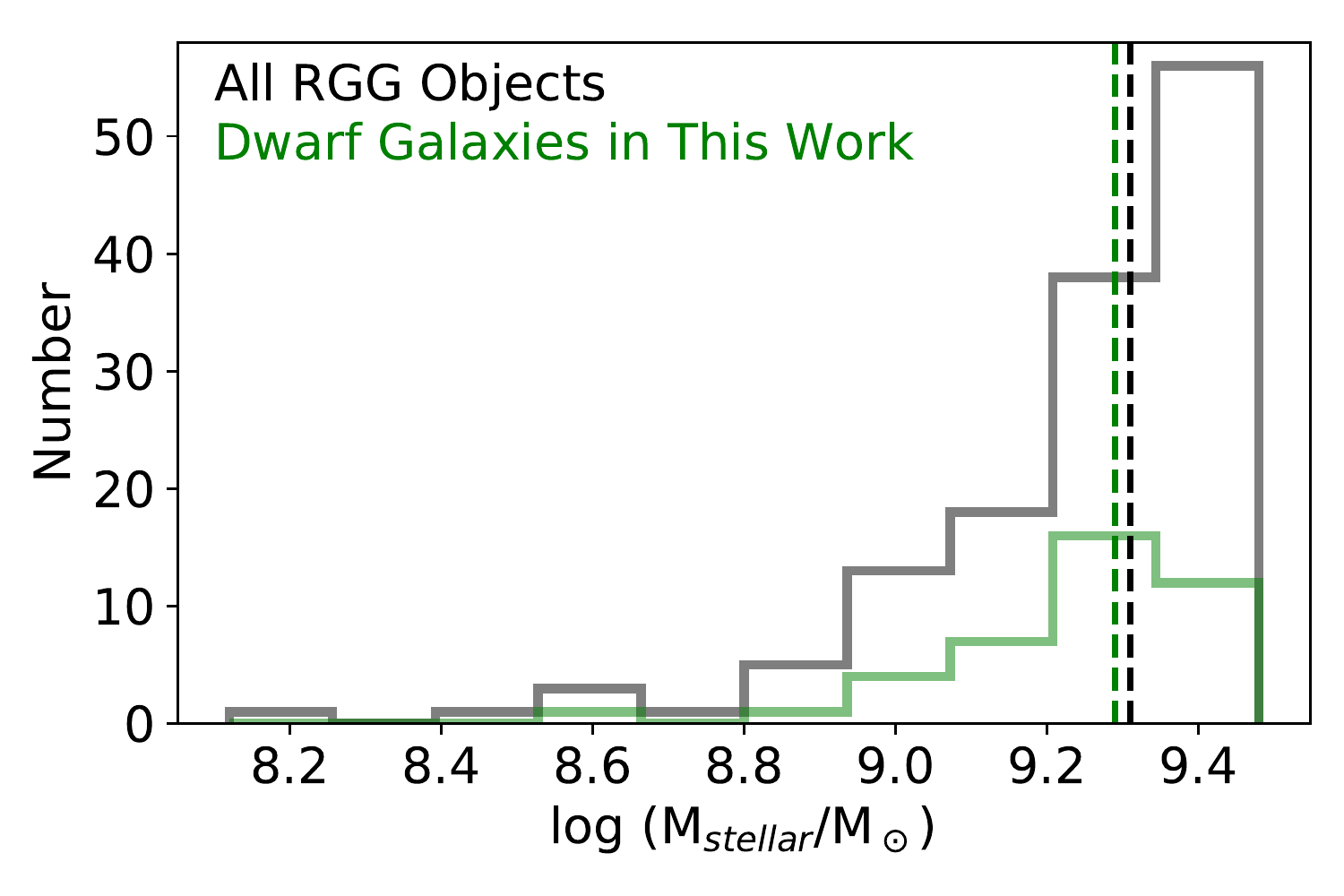}} &
{\includegraphics[width=0.42\textwidth]{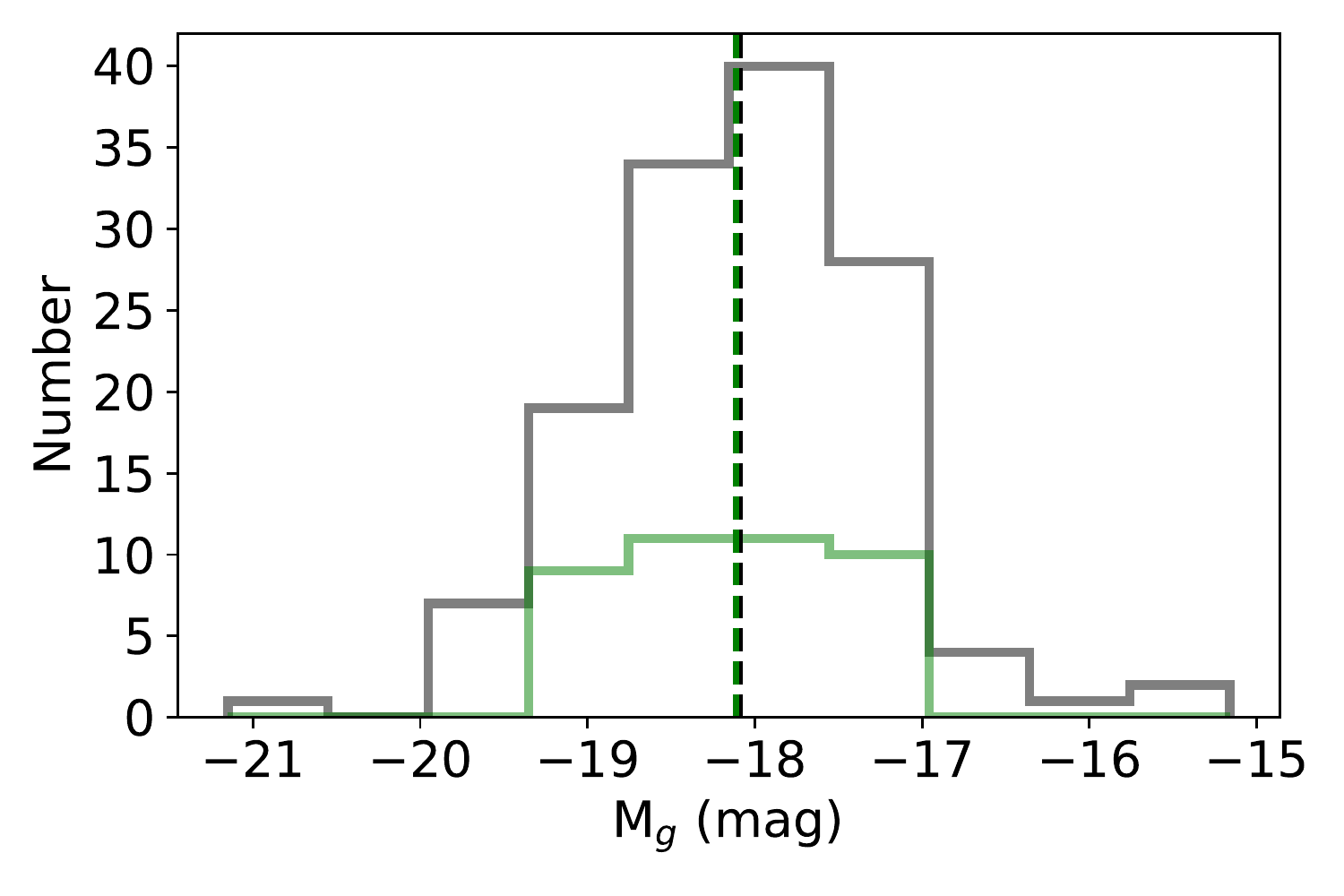}} \\
{\includegraphics[width=0.42\textwidth]{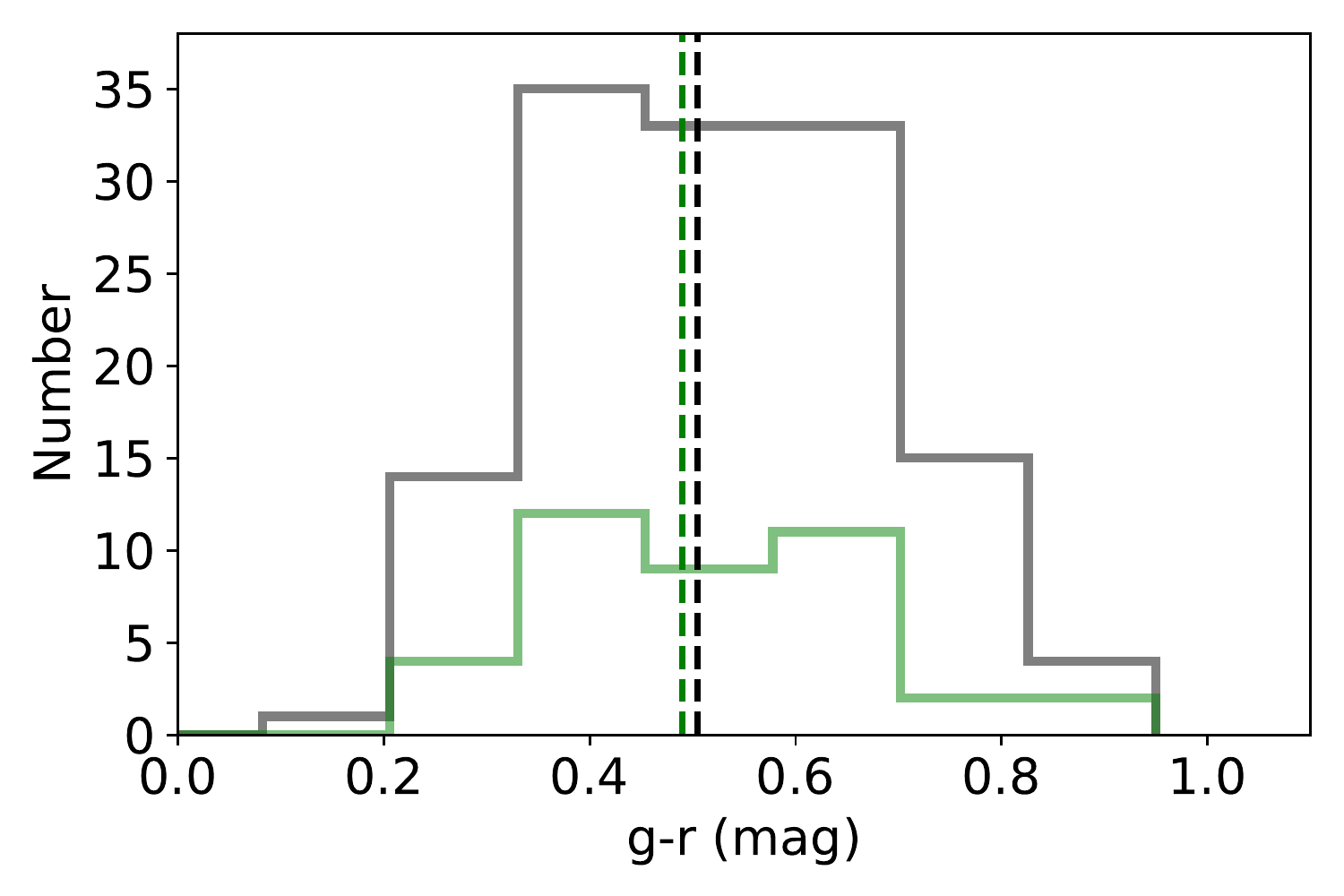}} &
{\includegraphics[width=0.42\textwidth]{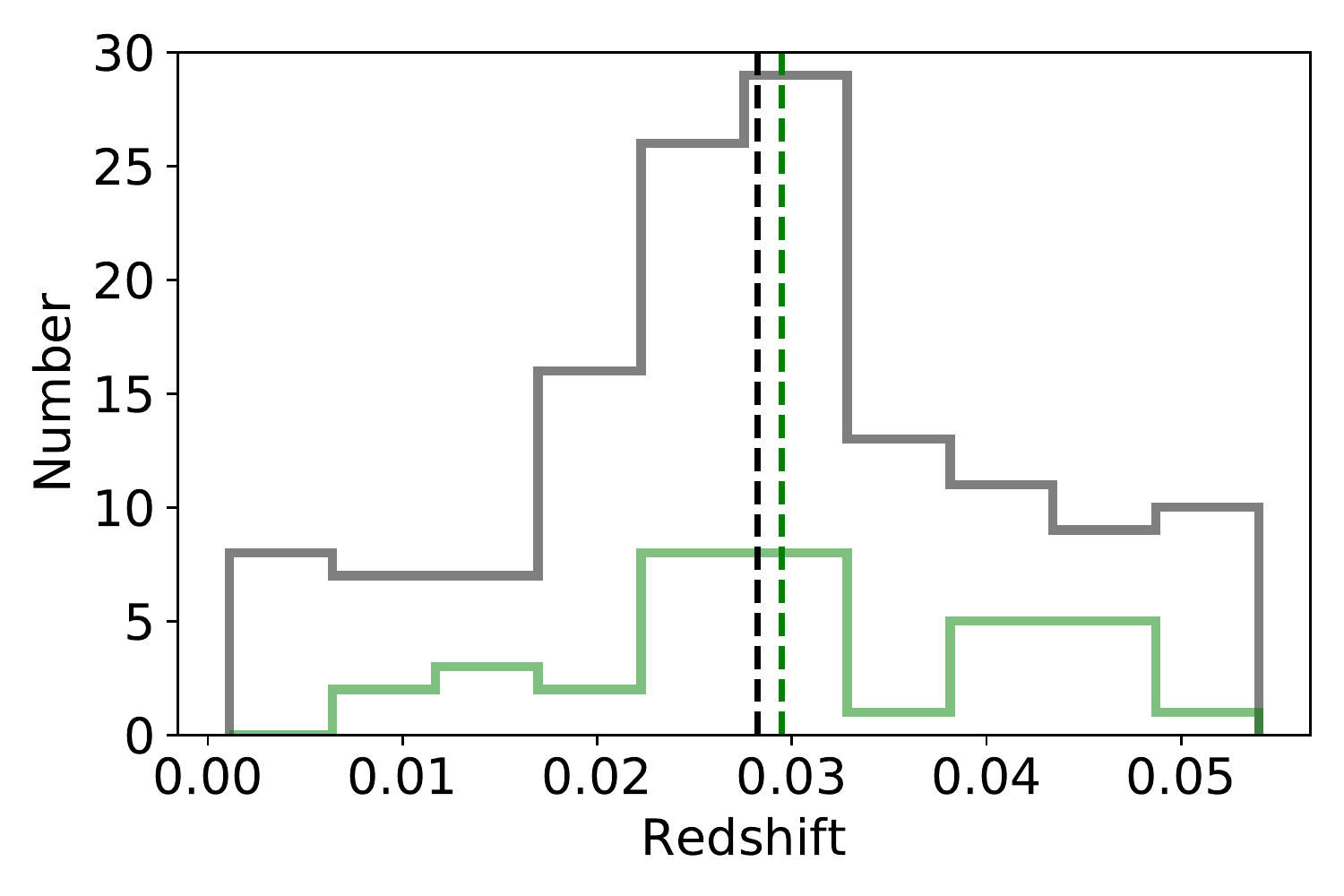}} \\
{\includegraphics[width=0.42\textwidth]{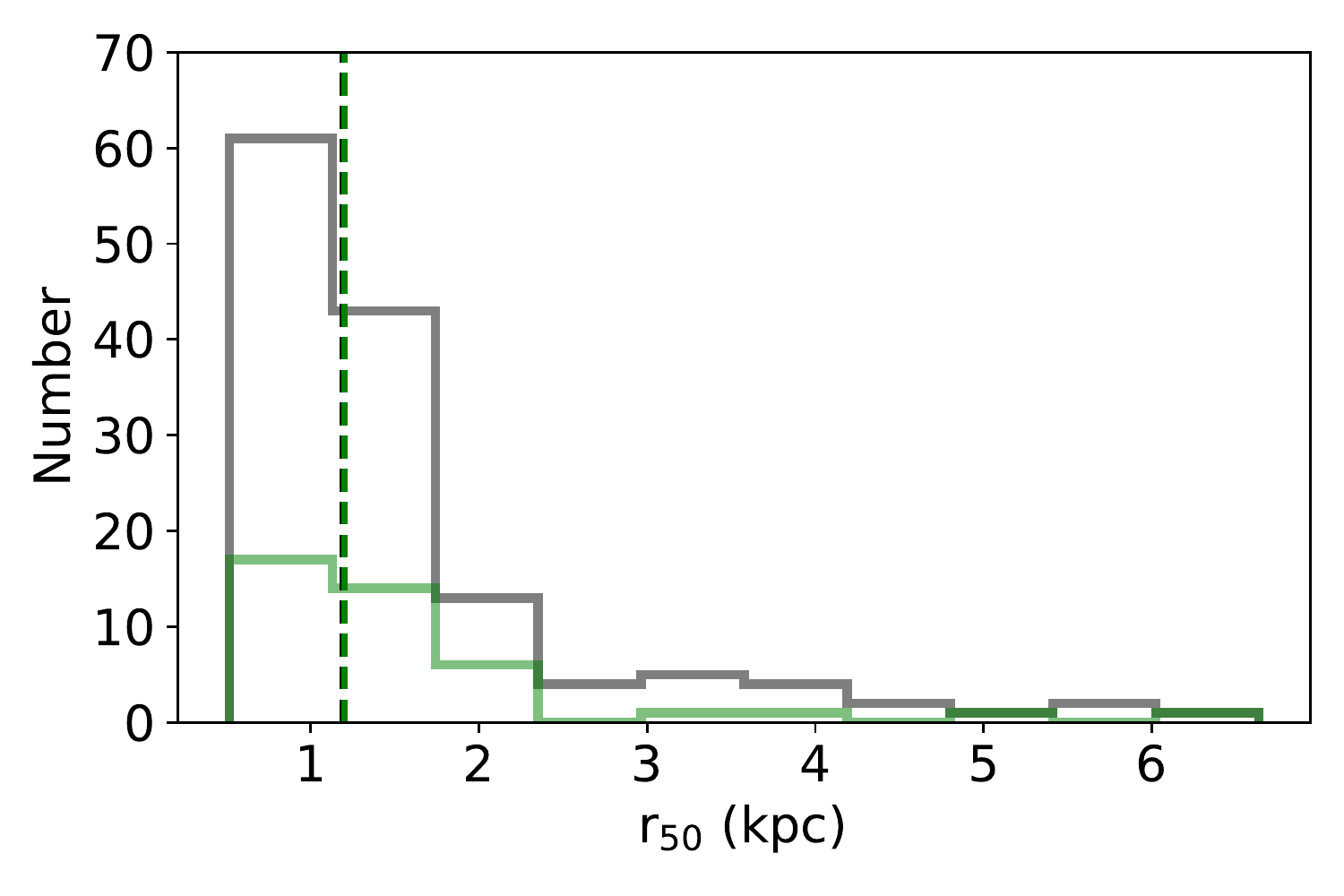}} &
{\includegraphics[width=0.42\textwidth]{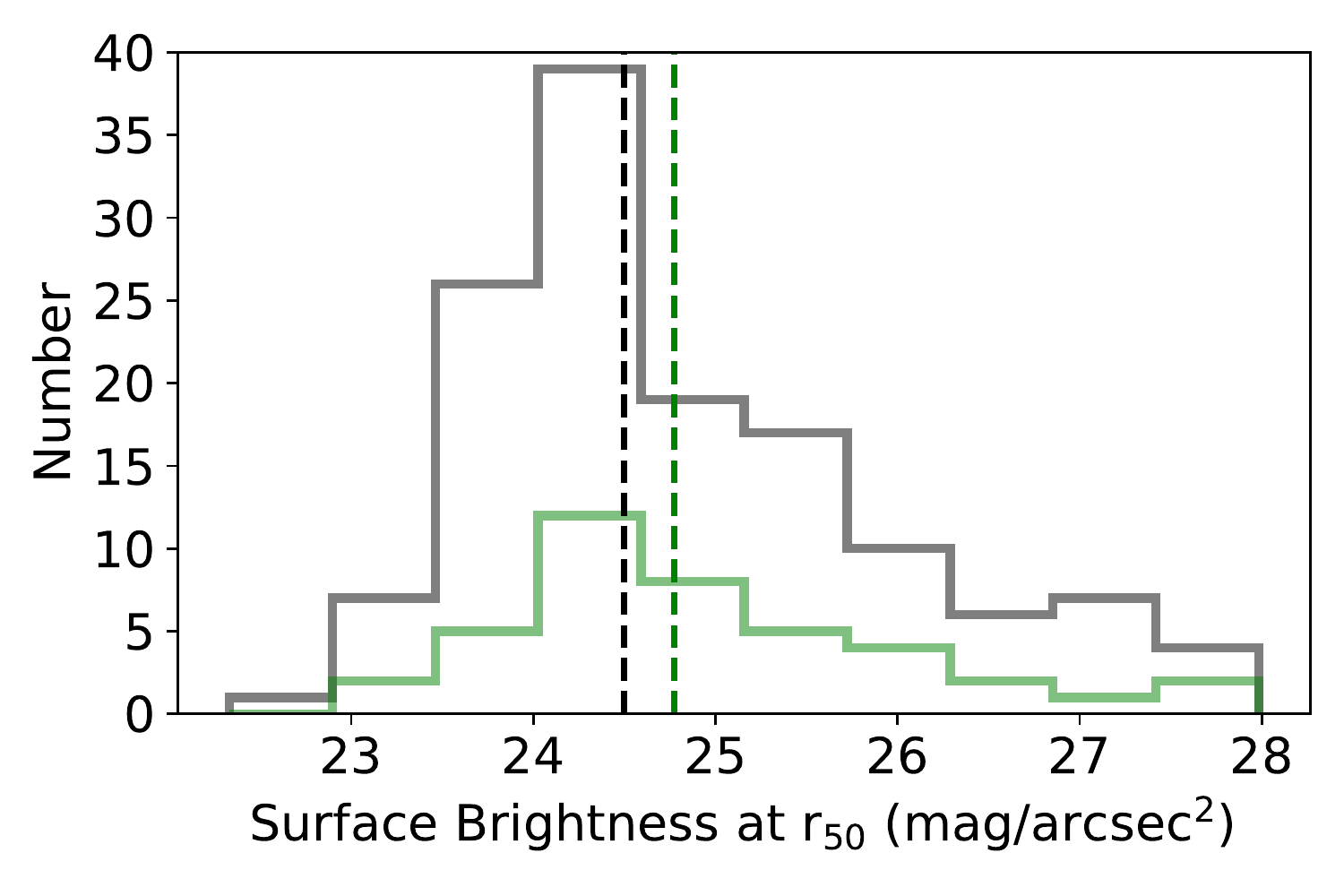}}
\end{array}$
\caption{Distribution of stellar mass, absolute $g$-band magnitude, $g-r$ color, redshift, Petrosian half-light radius and interpolated g-band surface brightness at the Petrosian half-light radius for the galaxies in this work (green histograms). Properties of the entire \citet{reines} (RGG) sample are shown as gray histograms. Dashed lines indicate median values.}
\label{fig:sampleproperties}
\end{center}
\end{figure*}

\section{Sample of Dwarf Galaxies}\label{sec:sample}

The dwarf galaxies with optically-selected AGNs studied here are a subset of objects identified by \citet{reines}. Starting with a sample of $\sim 25,000$ emission line galaxies with stellar masses $M_\star \leq 3 \times 10^9 M_\odot$ in the NASA-Sloan Atlas (NSA), \citet{reines} analyzed Sloan Digital Sky Survey (SDSS) spectra of these objects and found 136 dwarf galaxies exhibiting optical signatures of accreting massive BHs. These galaxies fall in either the AGN or Composite region of the [OIII]/H$\beta$ vs. [NII]/H$\alpha$ narrow emission line diagnostic diagram (i.e., the BPT diagram; \citet{baldwinetal1981, KewleyStarburst}). Our target galaxies were also selected to be in the Seyfert region of the [OIII]/H$\beta$ vs. [SII]/H$\alpha$ narrow emission line diagnostic diagram, making them among the strongest cases of dwarf galaxies hosting massive BHs. 
However, it is worth noting that low-metallicity AGNs and low-metallicity starbursts both fall in the upper left region of the BPT diagram and are therefore difficult to distinguish (i.e., RGG 5, the leftmost point in Figure \ref{fig:BPT}; also see the discussion in Section 6.3).

We proposed for {\it HST} SNAP observations of 61 dwarf galaxies meeting the criteria above and 33 were ultimately observed for this program. Of these, 13 galaxies were classified as AGNs by \citet{reines} and 20 were classified as Composites. Two of these galaxies, RGG 20 and RGG 118, are broad-line AGNs with virial BH masses of $M_{\rm BH} \sim 10^{6.1} M_\odot$ \citep{reines} and $M_{\rm BH} \sim 10^{4.7} M_\odot$ \citep{baldassare2015}, respectively. 

BH mass estimates for the narrow-line objects are in the range of $M_{\rm BH} \sim 10^{4.9} - 10^{5.8}M_\odot$ based on the AGN scaling relation between BH mass and total galaxy stellar mass from \citet{reinesvolonteri2015}. Total stellar masses from the NSA are in the range $M_\star \sim 10^{8.5} - 10^{9.5} M_\odot$, and come from the \texttt{kcorrect} code of \citet{blantonroweis2007}. 

We also include 7 additional dwarf galaxies with broad-line AGNs and Composites from \citet{reines} that have {\it HST} observations from another program (PI: Reines, Proposal ID 13943). Virial BH masses for these objects are in the range $M_{\rm BH} \sim 10^{4.9} - 10^{6.1} M_\odot$ \citep{reines} and these AGNs have been confirmed with {\it Chandra} X-ray observations \citep{baldassare2017xray}. The {\it HST} near-IR data are analyzed by \citet{schutte}, who also give an updated BH-bulge mass relation including dwarf galaxies. Here we include the galaxy structural information presented in \citet{schutte}. We also include 1 additional broad-line object (RGG 123) from \citet{reines} that has {\it HST} observations and structural analysis presented in \citet{jiang}, bringing our total sample to 41 BH-hosting dwarf galaxies with high quality {\it HST} images.

Figure \ref{fig:BPT} shows narrow-line diagnostic diagrams with locations of our sample galaxies using the line measurements in \citet{reines}. Figure \ref{fig:sampleproperties} shows the stellar masses, absolute magnitudes, colors, redshifts, half-light radii and g-band surface brightnesses of our galaxies compared to the entire \citet{reines} sample, which illustrates that we have a representative sample of the RGG galaxies in this work. In order to have uniform surface brightness measurements, we use data from the NSA giving surface brightnesses at discrete radii and use a spline interpolation to estimate the surface brightness at the Petrosian half-light radius for all the galaxies in the RGG sample, including the ones analyzed in this work. Table \ref{table:data} lists our sample of dwarf galaxies and their properties. 

\begin{deluxetable*}{lrccccc}
\tabletypesize{\footnotesize}
\tablewidth{\linewidth}
\tablecaption{Sample of 41 Dwarf Galaxies Hosting Optically-Selected AGNs with {\it HST} Observations \label{table:data}}
\tablehead{
\colhead{RGG ID} & \colhead{NSAID} & \colhead{SDSS Name} & \colhead{$z$} & \colhead{$g-r$ color} & \colhead{$M_g$} &\colhead{log $M_*$} \\
\colhead{(1)} & \colhead{(2)} & \colhead{(3)} & \colhead{(4)} & \colhead{(5)} & \colhead{(6)} & \colhead{(7)}
}
\startdata
\cutinhead{AGNs}
{\it RGG 1}$^a$ & 62996 & J024656.39$-$003304.8 & 0.0462 & 0.81 & $-$17.99 & 9.5 \\
RGG 2  & 7480    & J024825.26$-$002541.4 & 0.0247 & 0.58 & $-$17.32 & 9.1  \\
RGG 4   & 64339   & J081145.29+232825.7 & 0.0157 & 0.36 & $-$17.98 & 9.0 \\
RGG 5  & 46677   & J082334.84+031315.6 & 0.0098 & $-$0.29 & $-$18.85 & 8.5 \\
RGG 6  & 105376   & J084025.54+181858.9 & 0.0150 & 0.59 & $-$17.61 & 9.3 \\
RGG 7  & 30020   & J084204.92+403934.5 & 0.0293 & 0.62 & $-$17.45 & 9.3 \\
{\it RGG 9}$^a$ & 10779 & J090613.75+561015.5 & 0.0469 & 0.40 & $-$18.98 & 9.3 \\
RGG 10  & 106134   & J092129.98+213139.3 & 0.0313 & 0.58 & $-$18.20 & 9.3 \\
{\it RGG 11}$^a$ & 125318 & J095418.15+471725.1 & 0.0328 & 0.44 & $-$18.73 & 9.2 \\
RGG 15  & 27397   & J110912.37+612347.0 & 0.0068 & 0.36 & $-$17.33 & 8.9 \\
RGG 16  & 30370   & J111319.23+044425.1 & 0.0265 & 0.49 & $-$17.98 & 9.3 \\
{\it RGG 20}  & 52675   & J122342.82+581446.4 & 0.0144 & 0.66 & $-$18.11 & 9.5 \\
RGG 22  & 77431   & J130434.92+075505.0 & 0.0480 & 0.45 & $-$18.77 & 9.0 \\
RGG 26  & 54572   & J134939.36+420241.4 & 0.0411 & 0.47 & $-$18.55 & 9.3 \\
RGG 28  & 70907   & J140510.39+114616.9 & 0.0174 & 0.42 & $-$18.13 & 9.4 \\
RGG 29  & 71023   & J141208.47+102953.8 & 0.0326 & 0.42 & $-$17.74 & 9.1 \\ 
{\it RGG 32}$^a$ & 15235 & J144012.70+024743.5 & 0.0295 & 0.31 & $-$19.18 & 9.3 \\
\cutinhead{Composites}
RGG 37  & 6059    & J010005.93$-$011058.8 & 0.0514 & 0.59 & $-$18.76 & 9.3 \\
RGG 40  & 82616   & J074829.21+510052.4 & 0.0190 & 0.38 & $-$17.89 & 9.1 \\
{\it RGG 48}$^a$ & 47066 & J085125.81+393541.7 & 0.0411 & 0.28 & $-$19.19 & 9.1 \\
RGG 50  & 47918   & J090737.05+352828.4 & 0.0276 & 0.53 & $-$18.53 & 9.4 \\
RGG 53  & 105953   & J091720.88+191018.9 & 0.0285 & 0.87 & $-$17.17 & 9.3 \\
RGG 56  & 26850   & J093239.45+511542.9 & 0.0473 & 0.38 & $-$18.99 & 9.2 \\
RGG 58  & 39968   & J093821.54+063130.8 & 0.0224 & 0.48 & $-$18.05 & 9.4 \\
RGG 59  & 39954   & J094705.72+050159.8 & 0.0242 & 0.47 & $-$17.73 & 9.2 \\
RGG 64  & 106991   & J100423.33+231323.4 & 0.0266 & 0.49 & $-$17.53 & 9.1 \\
RGG 66  & 55081   & J101747.09+393207.7 & 0.0540 & 0.51 & $-$18.92 & 9.0 \\
RGG 67  & 12623   & J102149.12+635206.8 & 0.0211 & 0.62 & $-$16.99 & 9.0 \\
RGG 69  & 117416   & J102833.33+184513.9 & 0.0274 & 0.58 & $-$17.98 & 9.5 \\
RGG 79  & 19138   & J112957.62+653804.8 & 0.0439 & 0.56 & $-$18.21 & 9.5 \\
RGG 81  & 93958   & J113129.20+350958.9 & 0.0337 & 0.95 & $-$17.24 & 9.3 \\
RGG 86  & 66343   & J115359.06+130853.6 & 0.0226 & 0.67 & $-$17.59 & 9.3 \\
RGG 88  & 52494   & J115812.53+575322.1 & 0.0415 & 0.50 & $-$18.18 & 9.3 \\
RGG 89  & 32762   & J115922.33+511809.2 & 0.0297 & 0.78 & $-$17.32 & 9.3 \\
RGG 94  & 161692   & J122505.40+051945.9 & 0.0066 & 0.67 & $-$17.27 & 9.3 \\
{\it RGG 118}$^b$ & 166155 & J152303.80+114546.0 & 0.0243 & 0.37 & $-$18.38 & 9.4 \\
{\it RGG 119}$^a$ & 79874 & J152637.36+065941.6 & 0.0382 & 0.25 & $-$18.65 & 9.4 \\
{\it RGG 123}$^c$ & 18913 & J153425.58+040806.6 & 0.0395 & 0.35 & $-$18.16 & 9.1 \\
{\it RGG 127}$^a$ & 99052 & J160531.84+174826.1 & 0.0317 & 0.61 & $-$17.46 & 9.4 \\
RGG 135  & 4308    & J173202.96+595855.0 & 0.0291 & 0.39 & $-$18.91 & 9.4 \\
RGG 136  & 5563    & J235609.14$-$002428.6 & 0.0256 & 0.31 & $-$18.68 & 9.2 \\
\vspace{-0.3cm}
\enddata
\tablecomments{Column 1: identification number given in \cite{reines}. Column 2: NSA identification number. Column 3: SDSS name. Column 4: redshift. Column 5: $g-r$ color from the NSA. Column 6: Absolute $g$-band magnitude from the NSA. Column 7: Log total stellar mass from the NSA in units of $M_\odot$. Italicized RGG IDs indicate broad-line AGNs, for which the virial mass has been determined using broad H$\alpha$ emission (\citealt{reines}; \citealt{baldassare2015} for RGG 118). 
\tablenotetext{a}{Structural analysis adopted from \citet{schutte}.} 
\tablenotetext{b}{Structural analysis adopted from \citet{baldassare}.}
\tablenotetext{c}{Structural analysis adopted from \citet{jiang}.}} 
\end{deluxetable*}

\section{Hubble Space Telescope Observations}

{\it HST} near-infrared images of the 33 dwarf galaxies observed for our SNAP program were obtained with the Wide Field Camera 3 (WFC3) between 2015 October 11 and 2017 June 18 (PI Reines; Proposal ID 14251). The observations were taken in the IR/F110W filter (wide $YJ$-band) with a central wavelength of 1.15 $\mu$m.  

Snapshot observations of each galaxy were taken with a total on-source exposure time of $\sim 16$ minutes. We utilized a four-point sub-pixel dither pattern and used a $512 \times 512$ pixel subarray to avoid buffer dumps. The subarray has a field of view of $65.5\arcsec \times 65.5\arcsec$ and all of our galaxies fall well within the array. 

The images were processed by the STScI data reduction pipeline using the AstroDrizzle routine. The final cleaned, combined and calibrated images have an angular resolution of $\sim 0\farcs13$ (FWHM), corresponding to a range in physical scales of $\sim 17 - 131$ pc, and a scale of $\sim 70$ pc at the median distance of our sample (111 Mpc).

\section{Analysis}\label{sec:analysis}

The primary goal of this work is to characterize the structures of dwarf galaxies hosting optically-selected AGNs. For galaxies with regular morphologies, our general approach is to model each galaxy with a PSF plus either one or two S{\'e}rsic components for the galaxy light. For the galaxies requiring two S{\'e}rsic components, these structures can be ascribed to an inner bulge/pseudobulge component plus an outer disk component. In general, we do not attempt to model more complex structures, such as spiral arms or tidal features, although we include bars in a few cases. In dwarf galaxies, these features tend to be very faint and difficult to model, despite being fairly obvious to the eye. In the subsections below, we describe the details of our modeling and analysis. 

Six of our galaxies are irregular in shape and fitting these galaxy images with axisymmetric models proved impractical. We present the dwarf irregular galaxies in Section \ref{sec:irregulars}.

\subsection{PSF Construction}

It is important to use an accurate PSF to model the detector response to a point source, given that these galaxies are selected as AGN hosts. A PSF which does not properly capture that response can lead to inaccurate modeling of the galaxy as a whole.

\begin{figure}[!h]
\begin{center}
\includegraphics[width=3.1in]{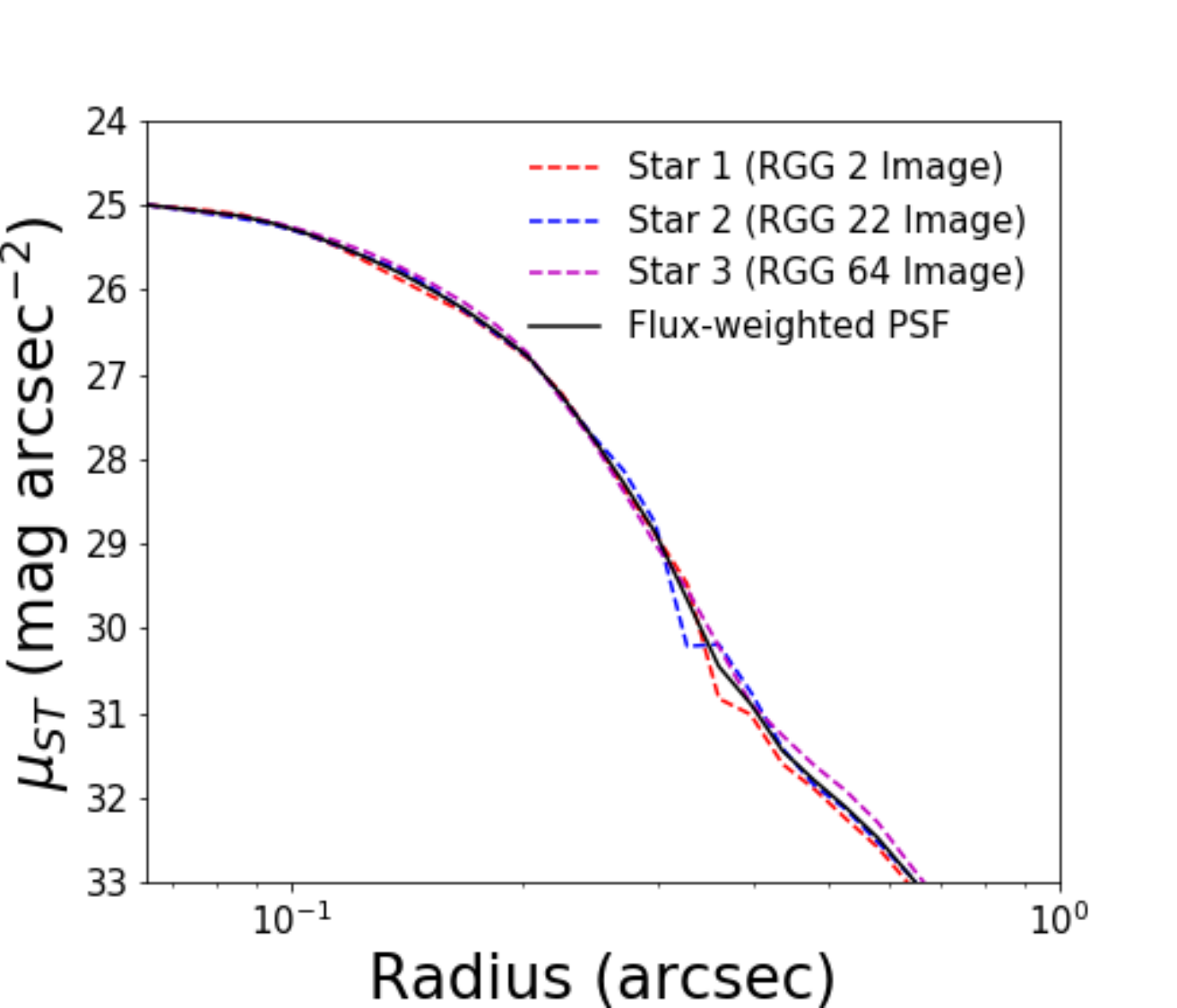}
\caption{Radial intensity profiles for the bright central stars in RGG 2, RGG 22 and RGG 64, as well as the flux weighted average PSF.}
\label{fig:psf}
\end{center}
\end{figure}

We used stars in our images to create PSFs. Since the response of a detector to a point source is dependent on location on the detector, we selected images with a bright star within 100 pixels of the center of the galaxy. Three stars met this criterion (one each in the images of RGG 2, RGG 22 and RGG 64).

We took a cutout of these stars and performed a flux-weighted average to create one image. We then made a model of the resulting image, giving us a PSF that is the result of averaging individual stars, while having the high signal-to-noise ratio required of PSFs. The PSF we created, as well as the profiles of each of the three stars, can be seen in Figure \ref{fig:psf}. We used this PSF for all of our modeling.

\begin{figure*}[!t]
\begin{center}
{\includegraphics[width=.95\textwidth]{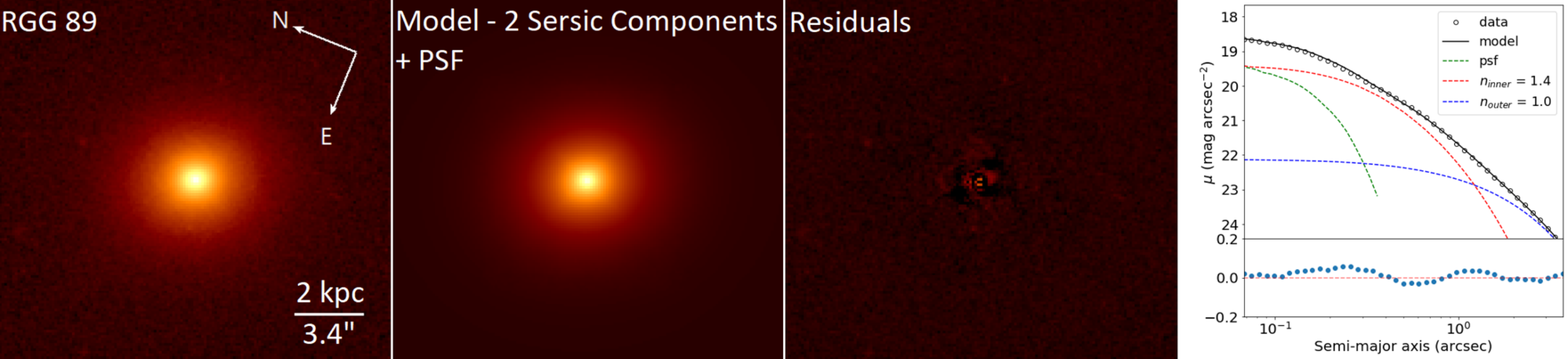}}
\caption{Left three panels: {\it HST}/WFC3 F110W image of RGG 89, GALFIT model, and the residuals after subtracting the model from the image. Images are shown on a stretched log scale to show faint details in the residuals. Right: Surface brightness profiles. The data are shown as black circles and the model is shown as a black line. The individual model components are shown as colored dashed lines. The residuals are shown in the bottom panel.}
\label{fig:RGG89}
\end{center}
\end{figure*}

\subsection{Galaxy Modeling}

We fit two-dimensional surface brightness models to our images using the galaxy fitting software GALFIT \citep{peng}. GALFIT has many analytical models which can be used to fit galaxies. For this work, we used the very general S{\'e}rsic profile, which takes the form \citep{sersic}:

\begin{equation}
  \Sigma(r) = \Sigma_e \times \textrm{exp}\bigg[-\kappa\Big(\Big(\frac{r}{r_e}\Big)^{\frac{1}{n}}-1\Big)\bigg]
\end{equation}

\noindent
where $r_e$ is the effective radius defined such that half of the flux lies within $r_e$, and $\Sigma_e$ is the surface brightness at $r_e$. The parameter $n$, which is coupled to $\kappa$, is the S{\'e}rsic index; a higher S{\'e}rsic index indicates a steep inner brightness profile and more extended wings, while a lower index indicates a shallower profile at small radius and less extended wings. $\kappa$ is a free parameter defined as $\kappa = {\Gamma(2n)}/{2}$, where $\Gamma$ is the complete gamma function. The case of $n = 4$ is the de Vaucouleurs profile, often used to model classical bulges \citep{devauc}. Values of $n = 1$ and $n = 0.5$ describe exponential disks and Gaussian profiles, respectively.

Our galaxy fitting process followed that recommended by \citet{peng}. Before running GALFIT, we first created mask files containing the pixel coordinates of prominent foreground and background objects we wished to ignore during the modeling process. We then began by using a single S{\'e}rsic component only.

We found that, in every case, a single S{\'e}rsic component was a poor fit and left very bright residuals. However, this gave us a rough estimate for some basic properties such as size and axis ratio of a given galaxy. We then added a central PSF to the single S{\'e}rsic model, using the results of the previous fit as our starting parameters. This PSF could represent the AGN and/or an unresolved nuclear star cluster (see \S \ref{sec:nsc}).

We then attempted to model each galaxy with two S{\'e}rsic components without a central point source. We let the S{\'e}rsic component of the inner component vary, while holding the S{\'e}rsic index of the outer component fixed at the canonical value for an exponential disk of $n=1$. This again gave us some basic information about the bulge/disk model (e.g., relative sizes) but it rarely resulted in a good fit. We then added a point source to the two-S{\'e}rsic model.

For three of the galaxies in our SNAP program, RGG 7, RGG 29 and RGG 37, a bar was necessary. This was determined through visual inspection of the data, in which the residuals showed a bright bar through the center of the galaxy that was missed by our basic model. RGG 127 also required a bar \citep{schutte}.

In general, when GALFIT found a best-fit model, we re-fit the galaxy several times while varying the starting parameters to ensure that the software was not just falling into a local minimum of goodness of fit.

\subsection{Model Selection}\label{sec:model}

Next we determined whether each galaxy was best fit by one S{\'e}rsic component or two S{\'e}rsic components (with one being an exponential disk), plus a PSF. To decide this, we followed the example of \citet{oh2017} and used a three-step model selection process. First, we eliminated the two S{\'e}rsic model if the effective radius of the exponential disk was smaller than the effective radius of the inner component. Next, we eliminated the two S{\'e}rsic model if the disk was subdominant everywhere in the radial profile. Finally, we used the Akaike Information Criterion (AIC) \citep{akaike}. Assuming normally distributed noise, the AIC is calculated from $\chi ^2$ as:

\begin{equation}
{\rm AIC} = \chi ^2 + 2k
\end{equation}

\noindent where $k$ is the number of free parameters in a model. Following \citet{oh2017}, we eliminated the two S{\'e}rsic model if adding more parameters did not reduce the AIC by $\geq$ 10. We report the results in \S\ref{sec:results}. 

We note that the faintest features we modeled reach a surface brightness of $\sim$ 25 mag/arcsec$^2$. It is possible that, for some of the least luminous galaxies in our sample, {\it HST} could fail to detect features fainter than this in the diffuse outer regions of the galaxy. This could have an impact on the modeling procedure and model selection described above.

\subsection{Testing PSF Necessity}\label{sec:psf}
We also tested whether the point sources are necessary components in our models, given that some are subdominant to the galactic components (e.g. RGG 69). We used the AIC as our first criterion; as in \S\ref{sec:model}, we rejected the model including a PSF if it does not reduce the AIC by $\geq$ 10. Using this criterion, all of the galaxies we modeled prefered the inclusion of a PSF.

As additional checks on the necessity of the PSFs, we also examined the S{\'e}rsic indices of the inner components and the residuals. \citet{kormendy} set an upper bound for the S{\'e}rsic index of dwarf ellipticals at n $\lesssim$ 4, and bounds for pseudobulges and classical bulges at n $\lesssim$ 2 and n $\gtrsim$ 2, respectively. Other studies performing {\it HST} photometry of dwarf galaxies and low-mass AGN hosts (e.g. \citealt{jiang, schutte, coma}) find (pseudo)bulges and dwarf ellipticals with n $\leq$ 4. Therefore, we were skeptical of models without a PSF if they led to an inner S{\'e}rsic index much larger than this (i.e., n $\geq$ 5). 

Finally, we visually examined the residuals after subtracting the GALFIT model from the data. A S{\'e}rsic component attempting to account for a missing PSF leads to telltale rings in the residuals, alternating bright and dark, in the center of the galaxy. Therefore, we also rejected models without PSFs if they led to these rings.

To summarize, of the 26 regular galaxies modeled in our SNAP sample, all 26 preferred the model which includes a PSF based on the AIC. In addition, 9 have S{\'e}rsic indices n $\geq$ 5 when a PSF is not included, and 21 have rings in the residuals when a PSF is not included (some have both a high S{\'e}rsic index and rings in the residuals). Only 3 out of the 26 galaxies (RGG 2, RGG 59 and RGG 88) preferred the PSF model solely based on the AIC. While the presence of a point source is less certain in these three galaxies, we nevertheless adopted the PSF model for consistency with the rest of the sample.

\subsection{Uncertainty Calculation}\label{s:uncertainty}
Uncertainties in the GALFIT parameters were found following the example of \citet{baldassare}. Magnitudes are most sensitive to changes in the sky background, while effective radii and S{\'e}rsic indices are most sensitive to the point spread function. Using sigma clipping, we iteratively subtracted points that were 3$\sigma$ above the median of each full image to estimate a sky background and, once GALFIT converged on fit parameters for a model, we replaced the fit sky background with the estimated one. We used the change in magnitudes as our error. To determine the uncertainty in effective radii and S{\'e}rsic index, we replaced the PSF constructed from averaging three central stars with one constructed from a single bright central star. The changes in effective radius and S{\'e}rsic index were used as our error. 

\subsection{Surface Brightness Profiles}

\begin{figure*}[!t]
\begin{center}
$\begin{array}{ccc}
{\includegraphics[width=0.32\textwidth]{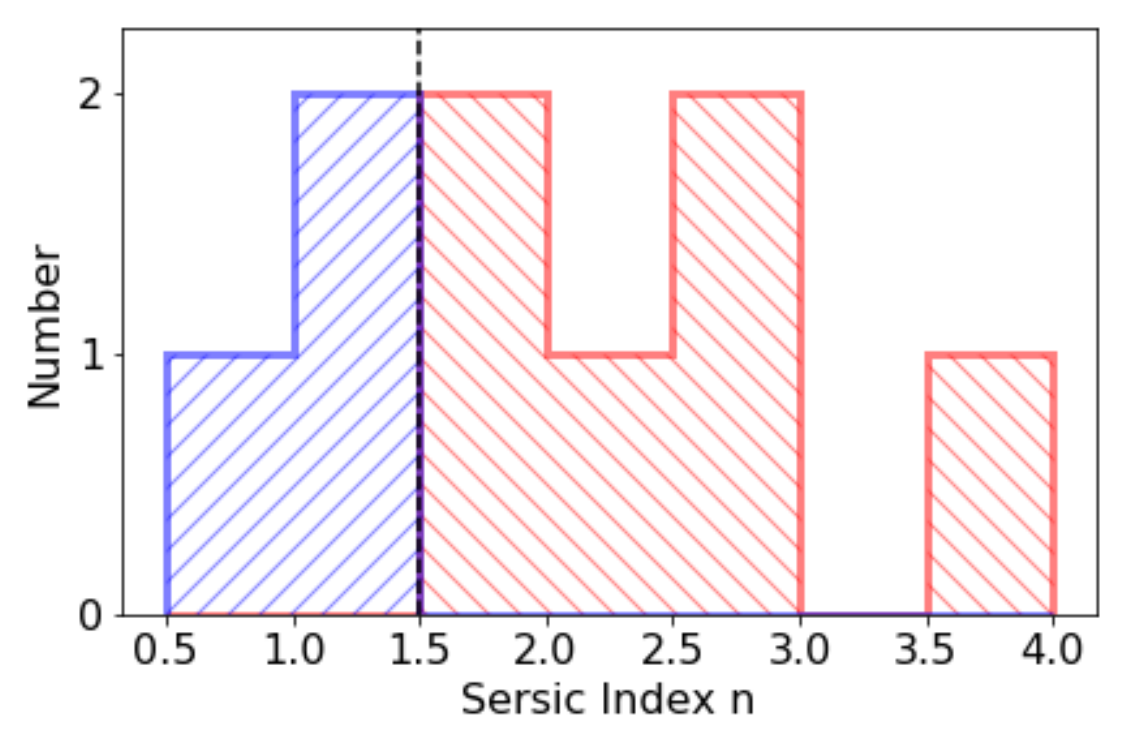}} &
{\includegraphics[width=0.32\textwidth]{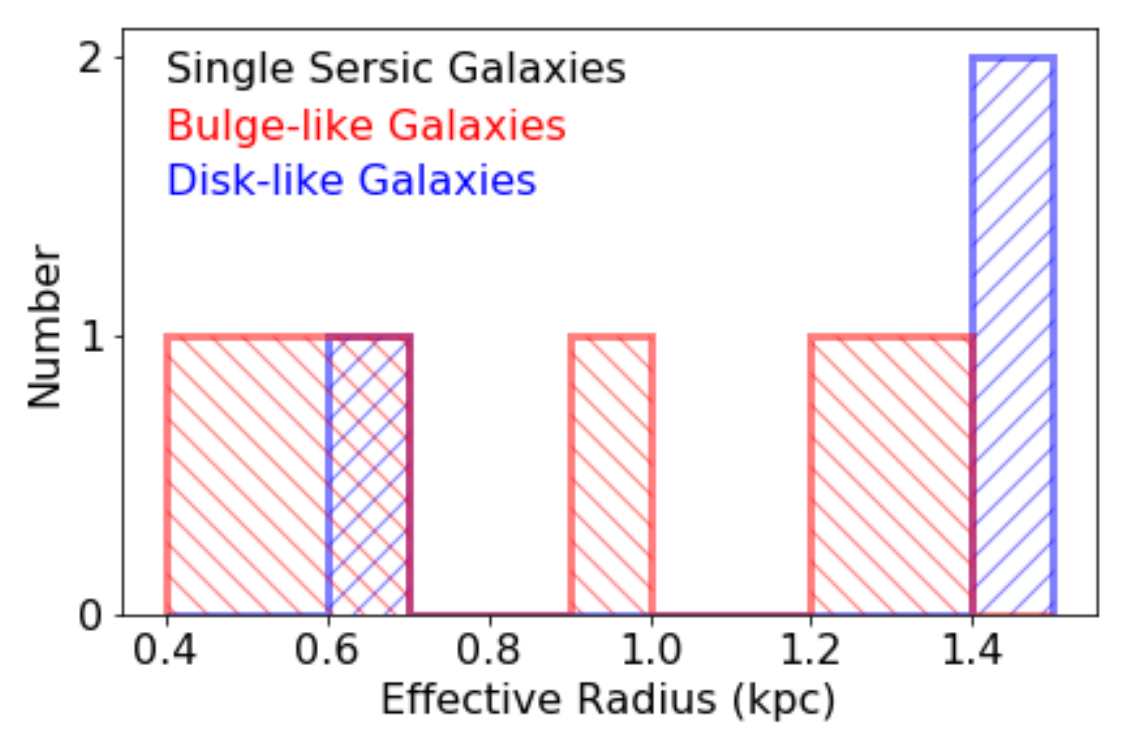}} & 
{\includegraphics[width=0.32\textwidth]{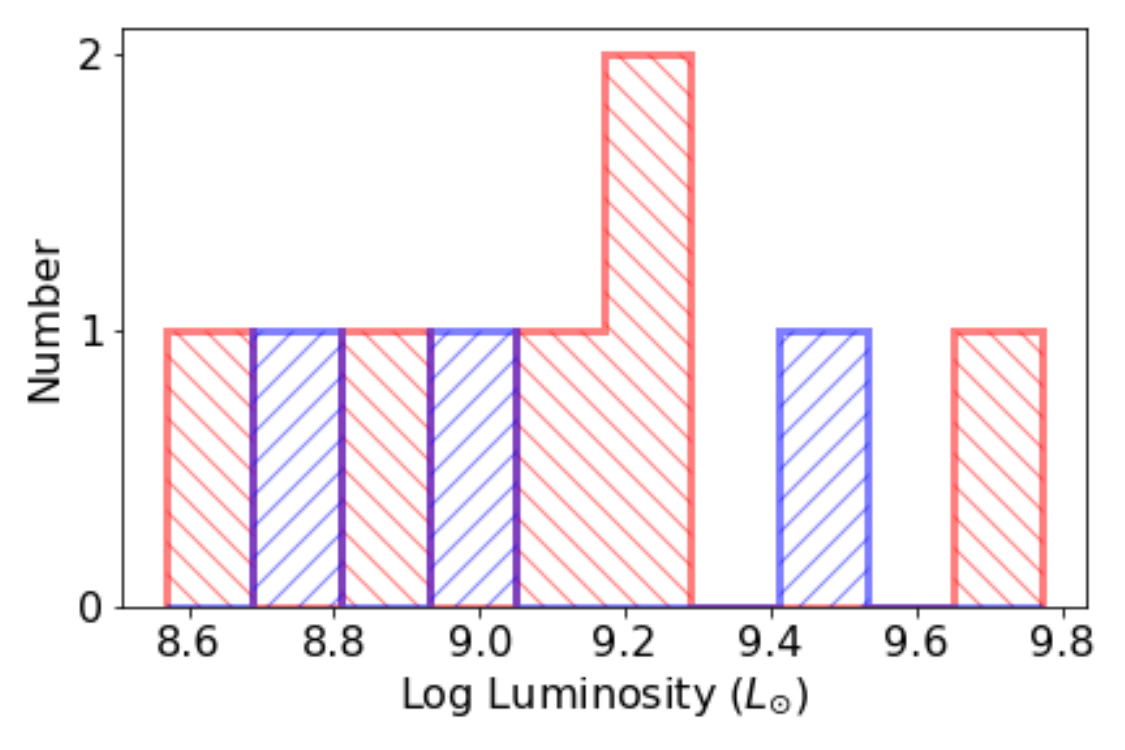}}
\end{array}$
\caption{Left: Distribution of the S{\'e}rsic indices for the one-S{\'e}rsic galaxies in our sample. A vertical dashed line shows the cutoff between disk-like galaxies and (pseudo)bulge-like galaxies. Disk-like galaxies are shown in blue histograms, while (pseudo)bulge-like galaxies are shown in red histograms. Middle: Distribution of the effective radii (in kpc), with the same color scheme as the left. Right: Distribution of the luminosities (in WFC3/IR F110W; 1.15$\mu$m), with the same color scheme as the left.}
\label{fig:soleprops}
\end{center}
\end{figure*}

Using the \textit{isophote} Python package, we fit elliptical isophotes to the data and our GALFIT models. With these isophotes, we constructed 1-D surface brightness profiles. We used these 1-D profiles as additional checks on our models. 
Figure \ref{fig:RGG89} shows an example of a GALFIT model and surface brightness profile for the dwarf galaxy RGG 89. Similar plots for the other galaxies in our sample are shown in the Appendix. The images are stretched on a log scale to show very faint details. While there are features in the residuals for some of the galaxies, they are quite dim and will not significantly affect our results. Typically, the residuals are $\lesssim 0.1$ mag arcsec$^{-2}$.

\section{Results}\label{sec:results}

Of the 41 AGN-hosting dwarf galaxies in our sample, the vast majority (35/41 = 85\%) have fairly regular morphologies and can be adequately modeled in GALFIT using one or two S{\'e}rsic components plus a PSF, and sometimes a bar (\S\ref{sec:regulars}). A smaller, but non-negligible, fraction of the dwarf galaxies in our sample (6/41 = 15\%) have irregular/disturbed morphologies and were not successfully modeled in GALFIT. These galaxies are presented in \S\ref{sec:irregulars}.

\subsection{Dwarf Galaxies With Regular Morphologies}\label{sec:regulars}

We have determined whether a single S{\'e}rsic or two S{\'e}rsic model is a better fit for the regular galaxies in our sample 
(including the galaxies taken from the literature as described in \S\ref{sec:sample} and noted in Table 1). A PSF component is required in all cases to account for the central AGN light and possibly an unresolved nuclear star cluster. 
The GALFIT results are summarized in the Appendix in Table \ref{table:secureresults}. An example {\it HST} image, model fit and surface brightness profile is shown in Figure \ref{fig:RGG89}. Corresponding figures for the other galaxies with regular morphologies can be found in the Appendix here (or see \citealt{schutte}, \citealt{baldassare}, \citealt{jiang}.)

We find that 74\% (26/35) of the regular galaxies in our sample are best fit by a two-component S{\'e}rsic model where we ascribe the outer component to a disk 
with fixed $n=1$.
The other 26\% (9/35) of the regular galaxies in our sample are best fit with a single S{\'e}rsic model.
 
For these single S{\'e}rsic galaxies, we distinguish between (pseudo)bulge/elliptical galaxies and disk-like galaxies using the S{\'e}rsic index; galaxies with $n \geq$ 1.5 are designated (pseudo)bulges/ellipticals, while galaxies with $n < 1.5$ are disk-like (see Figure \ref{fig:soleprops}).

\subsubsection{Disk Properties}\label{sec:disks}

\begin{figure}[!h]
\begin{center}
\includegraphics[width=0.4\textwidth]{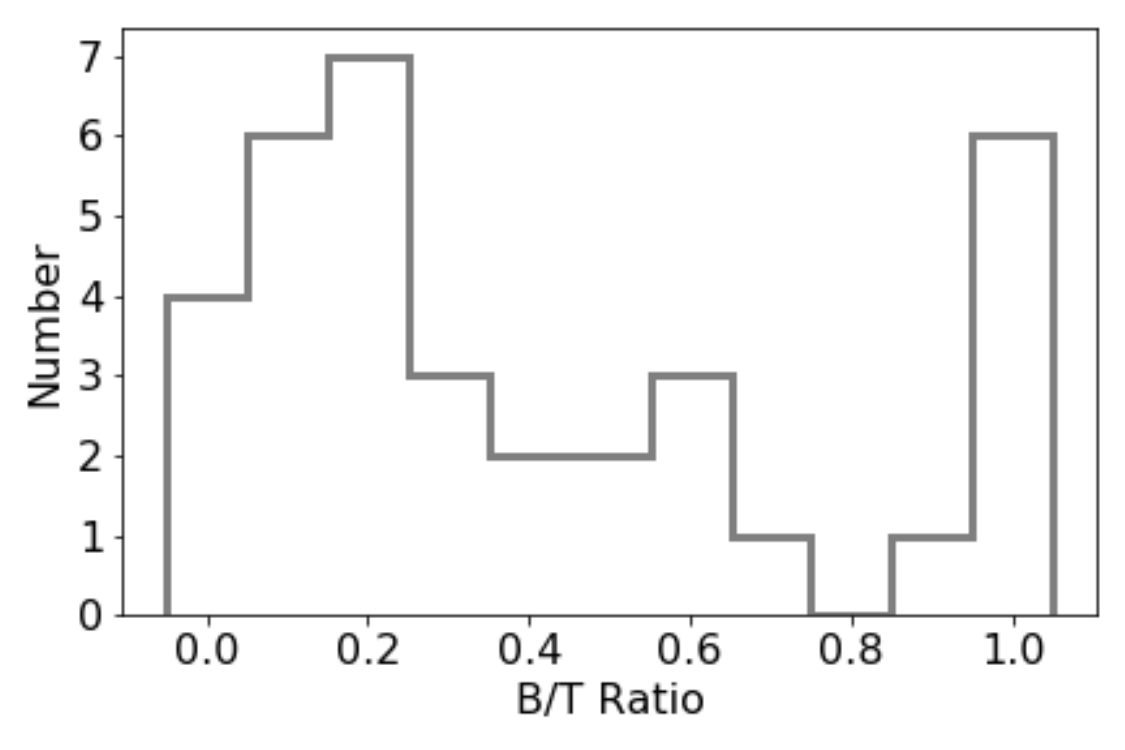}
\caption{Distribution of bulge-to-total light ratios for all the galaxies in our sample (one-S{\'e}rsic and two-S{\'e}rsic galaxies). A $B/T$ of 0 indicates a disk-like single-S{\'e}rsic galaxy, while a $B/T$ ratio of 1.0 indicates a (pseudo)bulge single-S{\'e}rsic galaxy.}
\label{fig:btratio}
\end{center}
\end{figure}

\begin{figure*}[!t]
\begin{center}
$\begin{array}{ccc}
{\includegraphics[width=0.315\textwidth]{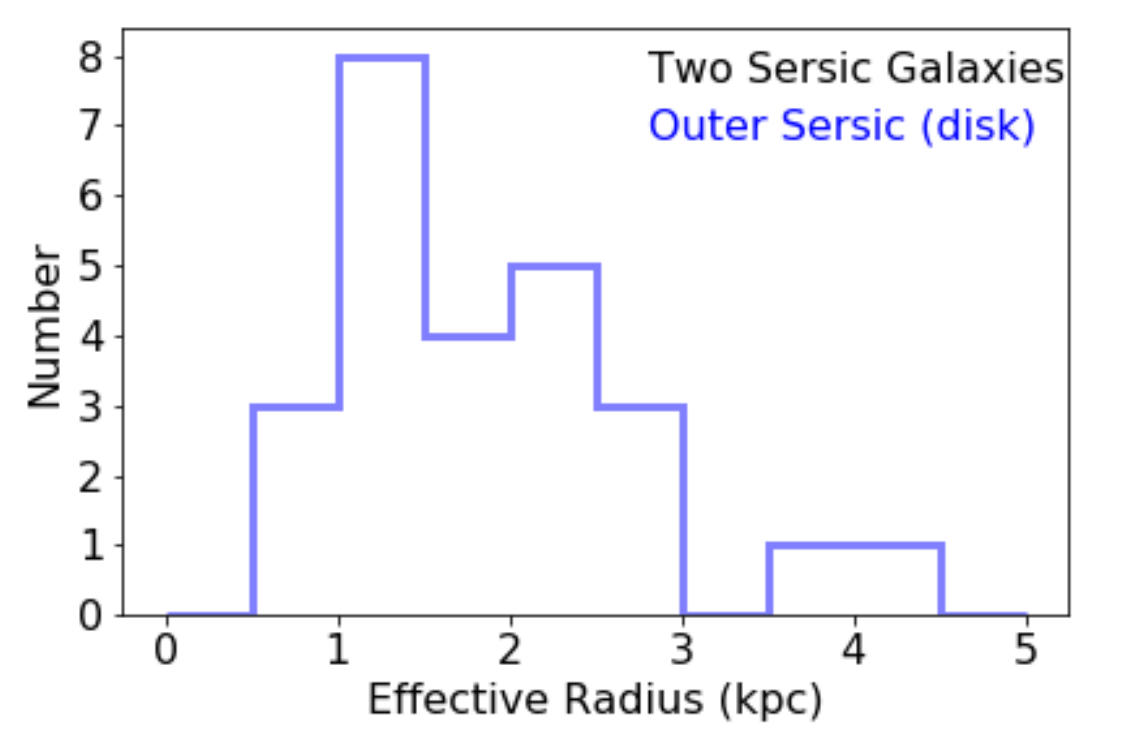}} &
{\includegraphics[width=0.315\textwidth]{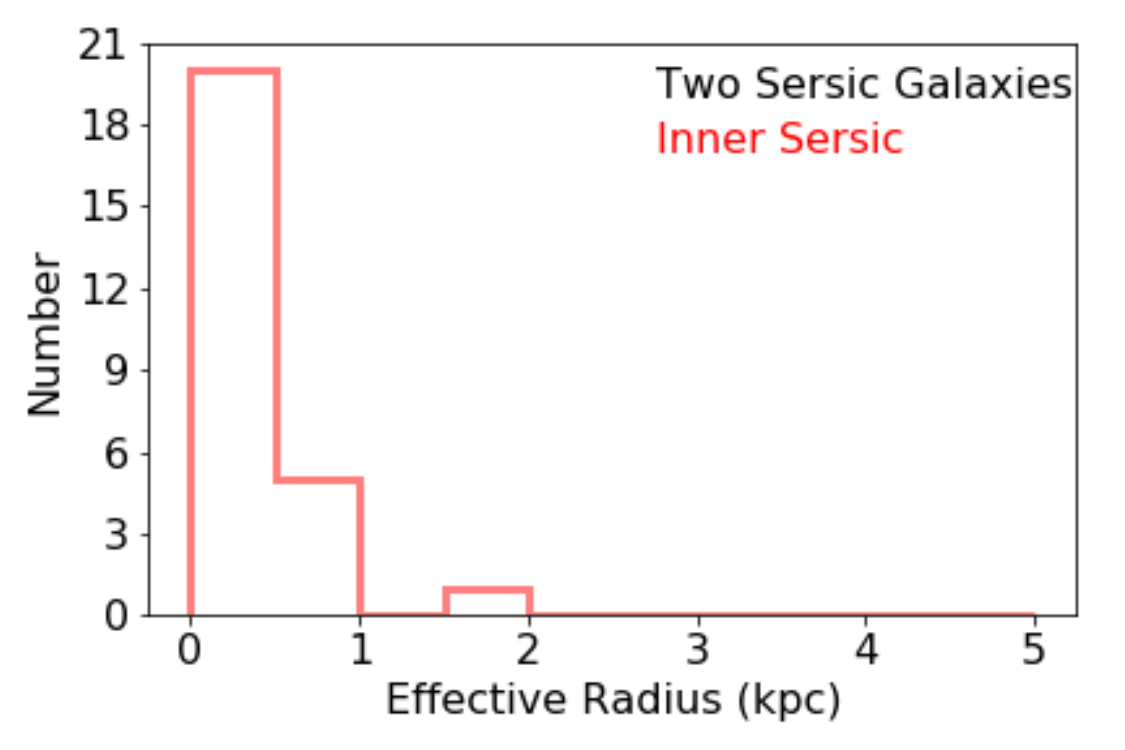}} & 
{\includegraphics[width=0.315\textwidth]{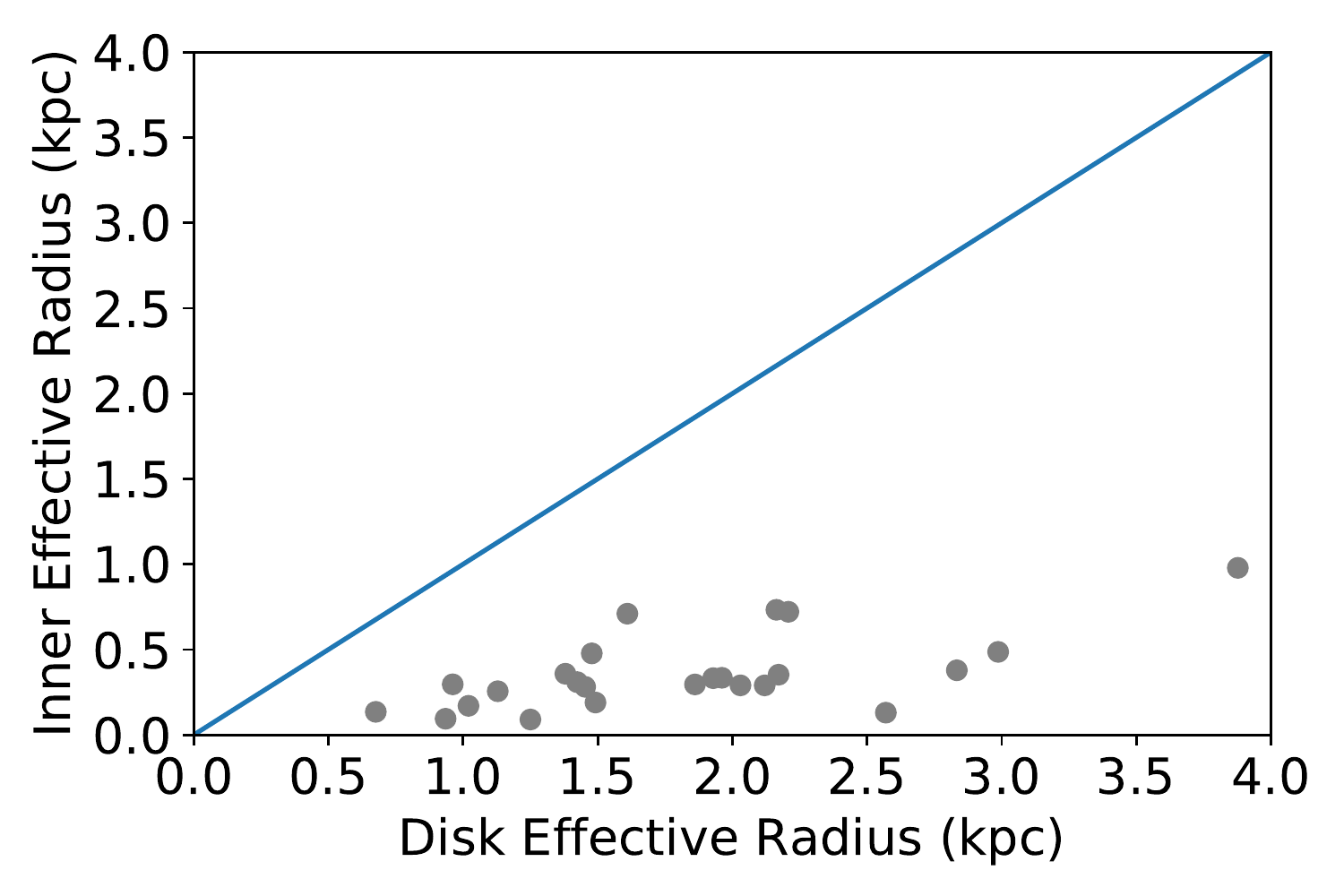}} \\
{\includegraphics[width=0.315\textwidth]{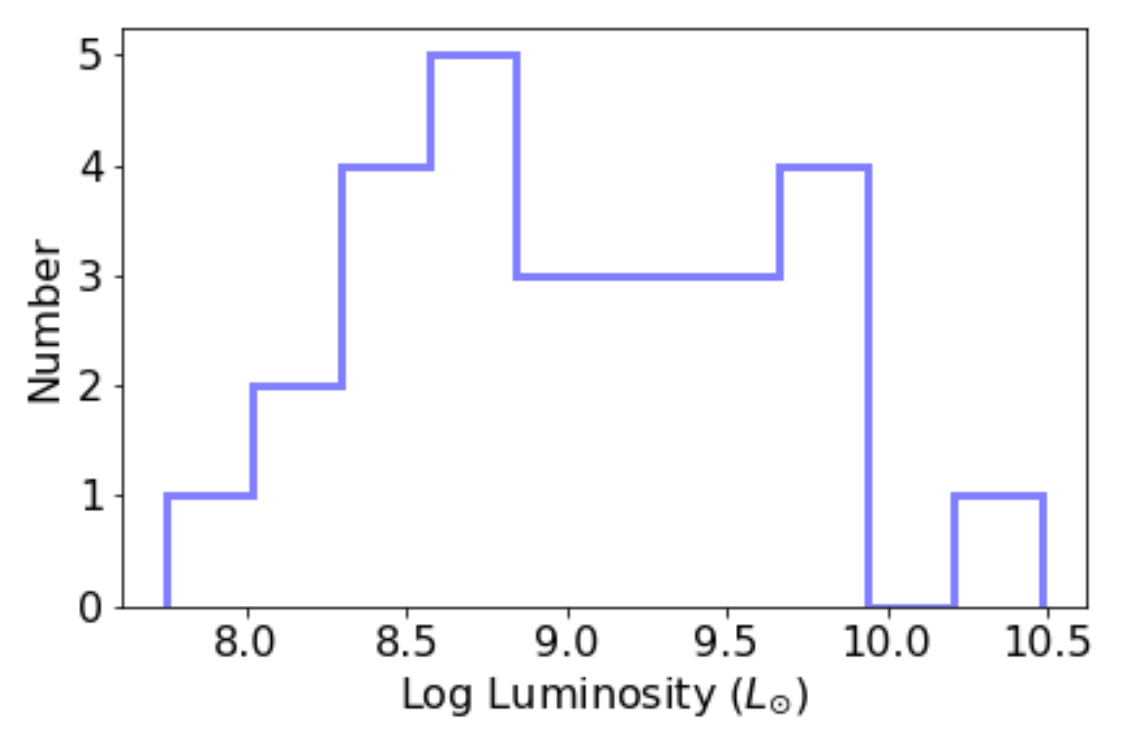}} &
{\includegraphics[width=0.315\textwidth]{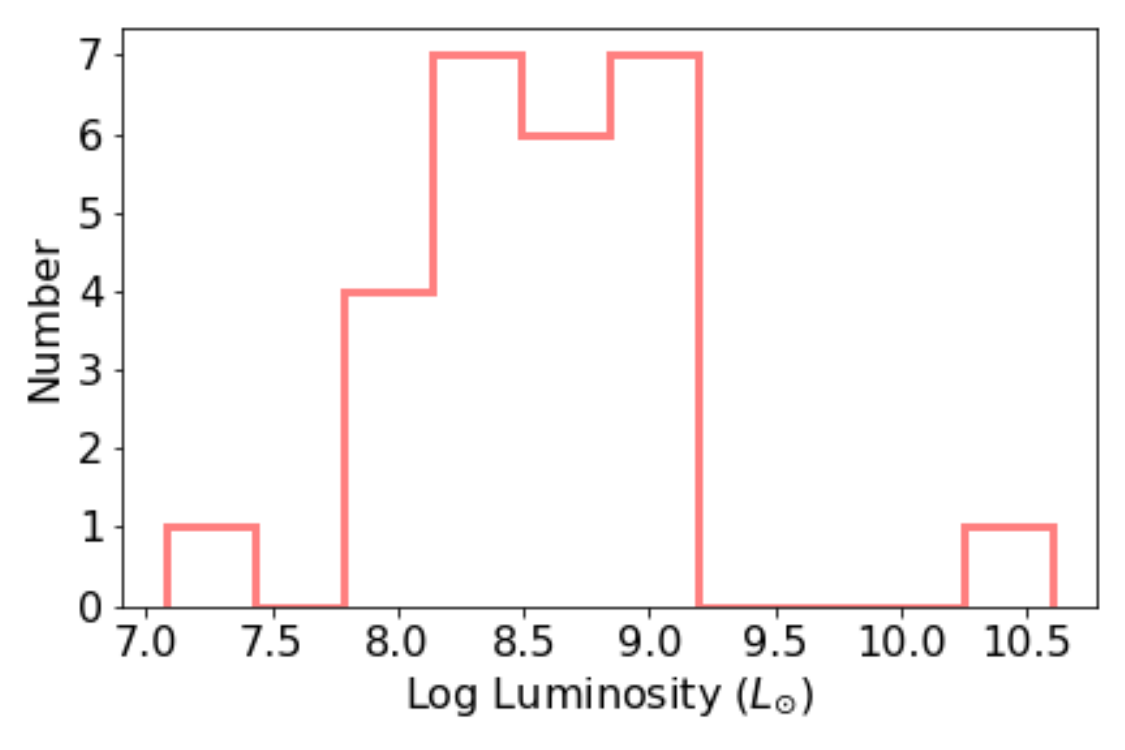}} & 
{\includegraphics[width=0.315\textwidth]{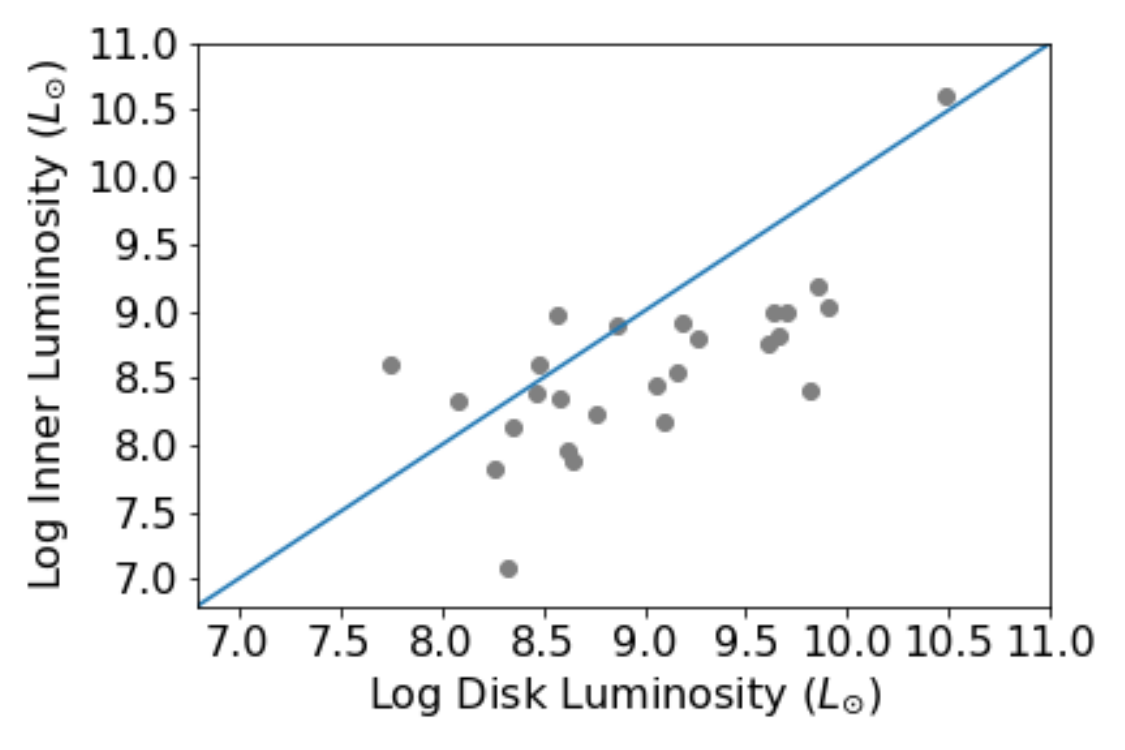}}
\end{array}$
\caption{Disk and (pseudo)bulge properties determined from GALFIT, for galaxies best fit with two S{\'e}rsic components. Left column from top to bottom: Disk effective radii (in kpc) and luminosities (in WFC3/IR F110W; $1.15 \mu$m). RGG 118, which has a disk effective radius of 6.5 kpc, is not shown in the left top panel. RGG 118 and RGG 123 are not included in the luminosity distribution in the lower left panel as the {\it HST} observations were taken in the WFC3/IR F160W and WFPC2 F814W filters, respectively. Middle column: Same as left column but for the inner S{\'e}rsic. Right column: (Pseudo)bulge versus disk properties for individual galaxies, along with the one-to-one line for comparison. Typical errors in the radii and luminosities are smaller than the points (see \S \ref{s:uncertainty} for discussion of error calculation and Table \ref{table:secureresults} for reported errors).}
\label{fig:prophists}
\end{center}
\end{figure*}

The galaxies in our sample tend to be disk-dominated. For the galaxies best fit with two S{\'e}rsic components, the median bulge-to-total $(B/T)$ ratio (with PSF subtracted) is $<{B/T}> = 0.21$ with a range of $B/T \sim 0.04 - 0.88$. Figure \ref{fig:btratio} shows the distribution of $B/T$ ratios. Galaxies modeled with a single S{\'e}rsic component consistent with a (pseudo)bulge are shown in Figure \ref{fig:btratio} as having a $B/T$ ratio of 1, and single S{\'e}rsic galaxies consistent with a disk have $B/T =0$. Luminosities of individual components were calculated from the apparent magnitudes reported by GALFIT and the distance to each galaxy. The half-light radii of the disks in our sample span a range of $r_{\rm e,disk} \sim 0.7 - 6.5$ kpc, with a median of $r_{\rm e,disk}\sim 2.2$ kpc. The left column of Figure \ref{fig:prophists} shows the distributions of $r_{\rm e,disk}$ and $L_{\rm disk}$. 

In addition to the 26 galaxies which have a disk and a (pseudo)bulge, 3 additional galaxies (RGG 7, 29 and 123) are best modeled with a single S{\'e}rsic component and appear to be disks without detectable (pseudo)bulges. However, it should be noted that RGG 7 and RGG 29 each have a bar in addition to the disk component, and RGG 7 also has spiral structure in the disk, which we do not model. 
There is also a bright point source at the center of these galaxies and it is possible we could have missed a very small bulge component (or nuclear star cluster), even with {\it HST} resolution. 

Additionally, RGG 15 is notable for being completely dominated by the disk at all radii except for the central PSF (see Figure \ref{fig:appendix2} in Appendix), although we do detect a small and faint inner S{\'e}rsic component with $n=0.5$. The half-light radius of the inner component is only $\sim 0.3$ kpc, compared to $\sim 2$ kpc for the disk. The inner component is also 3.5 magnitudes fainter than the disk of this galaxy.

\subsubsection{(Pseudo)bulge Properties}\label{sec:pseudobulges}

Given the importance of BH-to-bulge scaling relations (see, e.g., \citealt{kormendy}), it is vital to determine if our galaxies possess some kind of (pseudo)bulge component. In our sample of dwarf galaxies, we aim to distinguish between pseudobulges and classical bulges. Structurally, pseudobulges have more of an exponential profile than classical bulges, somewhat resembling an inner disk. The S{\'e}rsic index acts as an effective selector of pseudobulges versus classical bulges - pseudobulges have S{\'e}rsic indices $n \lesssim 2$ (see e.g., \citealt{fisher,kennicutt} for more on the differences between pseudobulges and classical bulges). 

Based on above criterion, 21 of the 26 regular galaxies best fit with two S{\'e}rsic components have an inner component consistent with a (pseudo)bulge, while 5 host a more classical bulge. 
 
When including the 6 single S{\'e}rsic galaxies that are consistent with being bulge-like, the total number of galaxies hosting a (pseudo)bulge comes to 32/41 (78\%). Despite most of our galaxies hosting some kind of (pseudo)bulge component, Figure \ref{fig:btratio} shows that they tend to be disk-dominated (\S\ref{sec:disks}).

\begin{figure}[!h]
\begin{center}
$\begin{array}{cc}
{\includegraphics[width=0.4\textwidth]{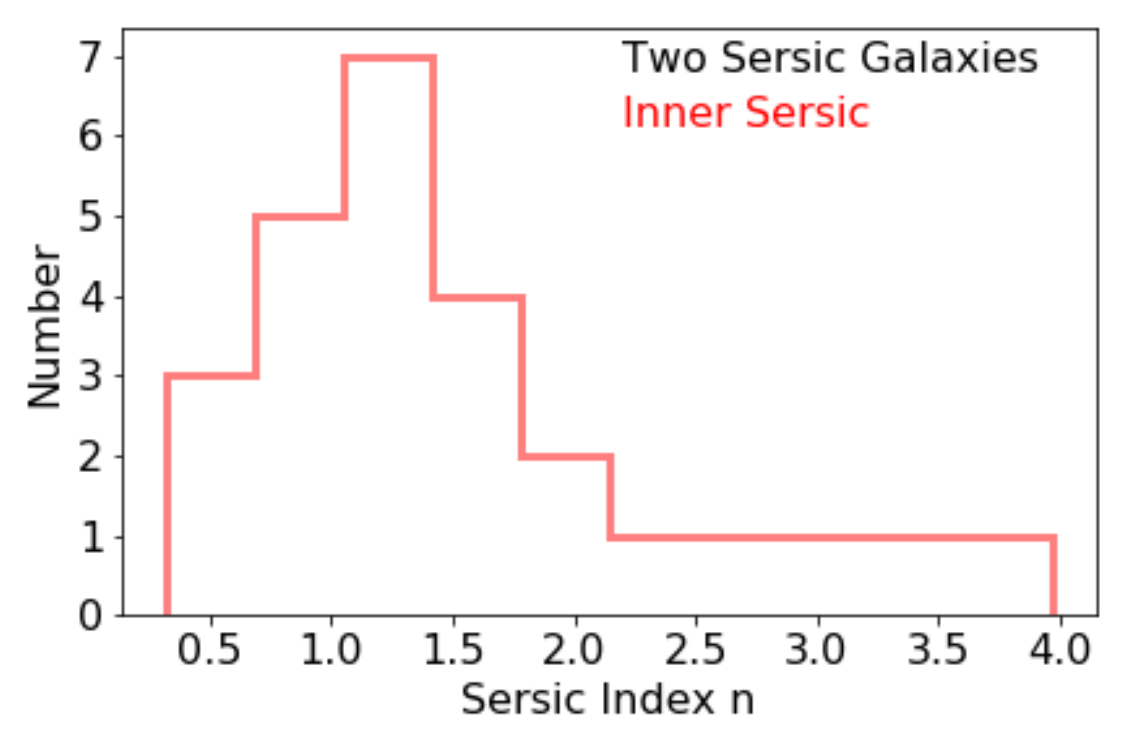}} \\
{\includegraphics[width=0.4\textwidth]{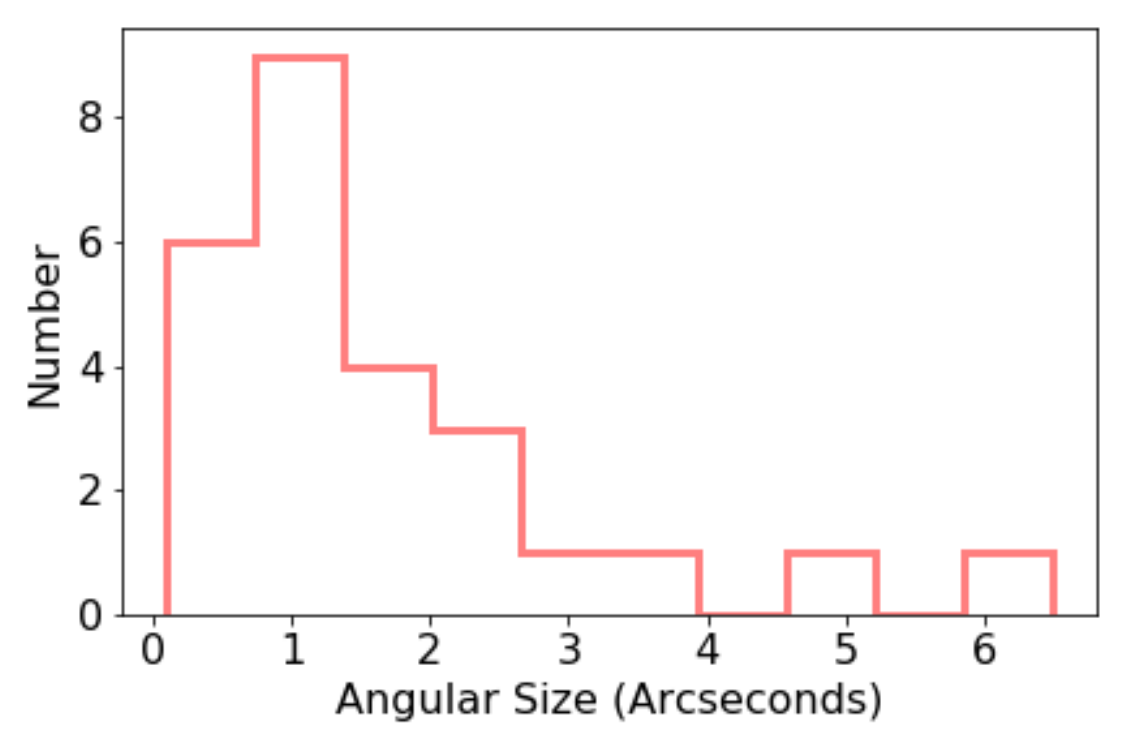}} & 
\end{array}$
\caption{Top: distribution of the S{\'e}rsic indices for the (pseudo)bulge components of our two-S{\'e}rsic galaxies. Bottom: distribution of the angular diameters of (pseudo)bulge components. We do not show the distribution of disk S{\'e}rsic indices since they are fixed to $n = 1$.} 

\label{fig:sersicsangularbt}
\end{center}
\end{figure}

The S{\'e}rsic indices of the inner/(pseudo)bulge components for the two-S{\'e}rsic galaxies are in the range of $n_{\rm bulge} \sim 0.3 - 4.0$, with a median of $<n_{\rm bulge}>=1.3$ (see Figure \ref{fig:sersicsangularbt}). The single S{\'e}rsic galaxies that are bulge-like have relatively large S{\'e}rsic indices compared to the inner components of galaxies best fit by a (pseudo)bulge--disk decomposition. 

The inner/(pseudo)bulge components in our dwarf galaxy sample tend to be quite compact. The half-light radii are in the range $r_{\rm e,bulge} \sim 0.1 - 1.6$ kpc, with a median of $<r_{\rm e,bulge}>\sim 0.3$ kpc. The most compact galaxy overall in our sample is RGG 29, a galaxy best fit with a single S{\'e}rsic component with $n=1.4$ and $r_{\rm e}\sim 0.5$ kpc. 

The middle column of Figure \ref{fig:prophists} shows the distributions of $r_{\rm e,bulge}$ and $L_{\rm bulge}$. 

\begin{figure*}[!t]
\begin{center}
\includegraphics[width=4.9in]{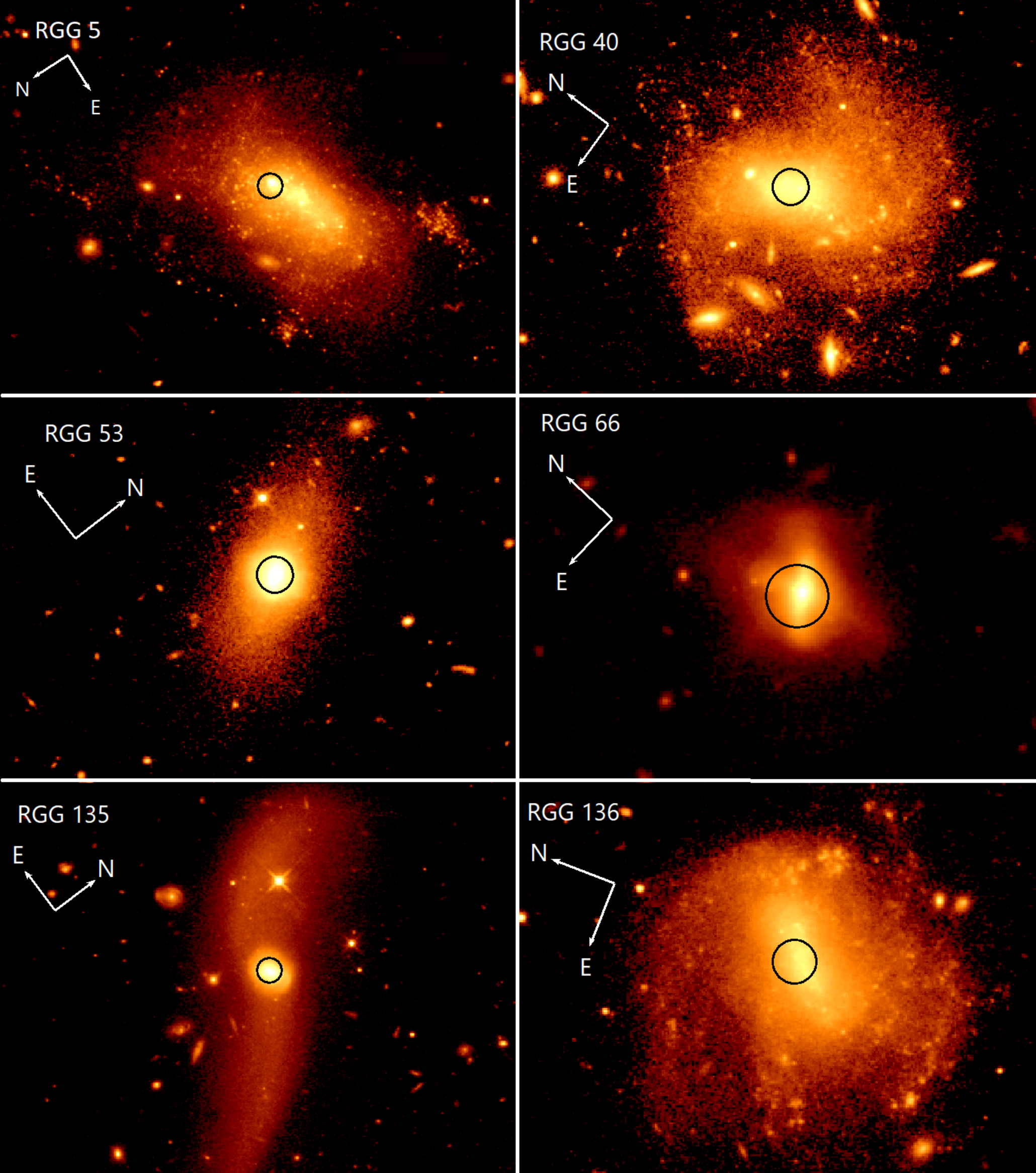}
\caption{{\it HST} observations of the dwarf irregular galaxies in our sample shown on a log scale. Black circles indicate the locations of the SDSS spectroscopic fibers, which are 3\arcsec\ in diameter. Fiber positions were retrieved from the NASA Sloan Atlas (NSA). While no astrometric corrections have been applied to the {\it HST} images, we checked that our astrometry is consistent with the Hubble Legacy Archive.}
\label{fig:irrimgs}
\end{center}
\end{figure*}

The angular diameters ($2r_{\rm e,bulge}$) of the (pseudo)bulges in our sample are shown in Figure \ref{fig:sersicsangularbt}. The median value is 1\farcs2 and the range in angular diameter is 0\farcs1 - 6\farcs5. This illustrates that {\it HST} resolution is essential to disentangle and detect small (pseudo)bulges in dwarf galaxies even at low redshift ($z \lesssim 0.05$).

\subsection{Irregular/Disturbed Dwarf Galaxies}\label{sec:irregulars}

Of the 41 dwarf galaxies in our sample, 6 (15\%) have irregular/disturbed morphologies and were not successfully modeled in GALFIT (RGG 5, 40, 53, 66, 135 and 136). The {\it HST} galaxy images are shown in Figure \ref{fig:irrimgs}. RGG 5, 40 and 136 do not have obvious photometric centers and resemble Magellanic-type dwarf irregulars. RGG 66 and 135 show signs of interactions/mergers, with RGG 135 displaying very elongated tidal tails. The remaining galaxy, RGG 53, looks more regular at first glance. However, a spiraled interior appears in the residuals when attempting to model the system in GALFIT. These galaxies are reminiscent of the dwarf galaxies found to host radio-selected AGNs by \citet{reines2020}, some of which host ``wandering" (i.e., non-nuclear) BHs.

With the exception of RGG 5, the irregular/disturbed dwarf galaxies all fall in the composite region of the [OIII]/H$\beta$ vs.\ [NII]/H$\alpha$ narrow-line diagnostic diagram. RGG 5 falls in the AGN part of the diagram just above the maximum starburst line; however, it is difficult to reliably distinguish between AGN and star-formation in this low-metallicity region of the diagram \citep{groves2006}. All of the galaxies were also selected to fall in the Seyfert region of the [OIII]/H$\beta$ vs.\ [SII]/H$\alpha$ diagram (see \S\ref{sec:sample}). Follow-up X-ray observations would help confirm if these irregular/disturbed dwarf galaxies do indeed host AGNs.

\begin{figure}[!h]
\includegraphics[width=0.45\textwidth]{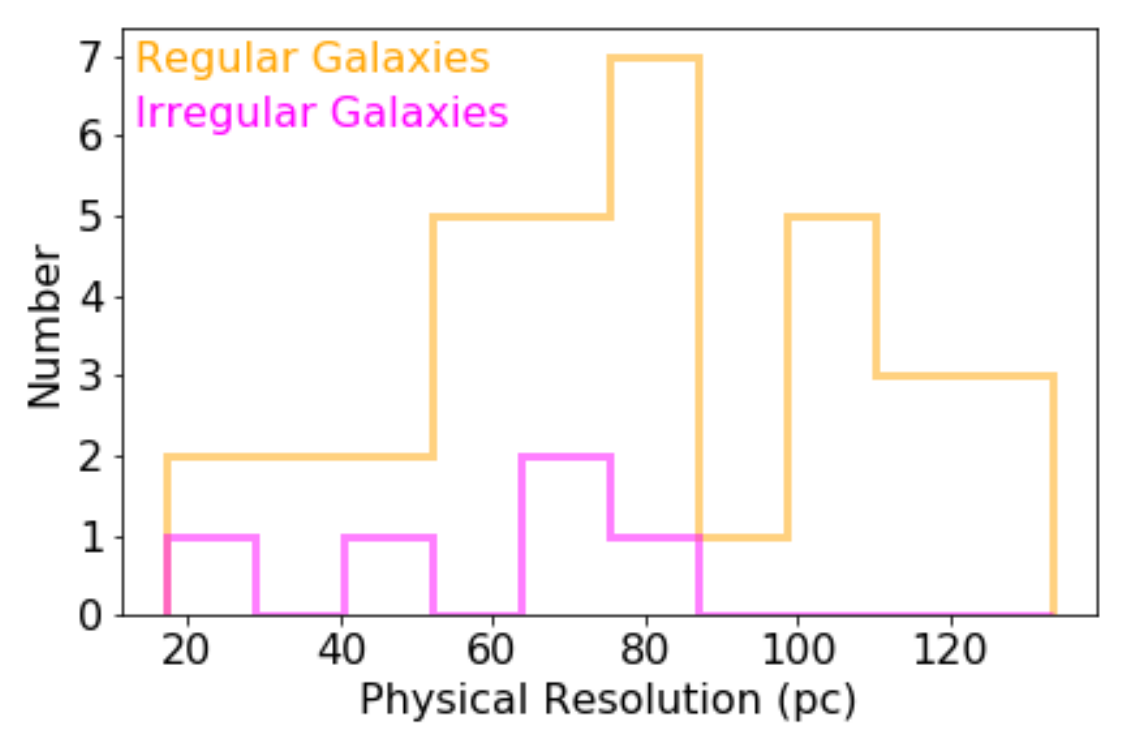}
\caption{Distribution of the physical resolution at the various distances of the galaxies in this work, with {\it HST} angular resolution (FWHM) of 0\farcs13. Regular galaxies are shown in orange histograms, while irregular galaxies are shown in magenta histograms.}
\label{fig:redshiftresolution}
\end{figure}

Figure \ref{fig:redshiftresolution} shows the distribution of physical resolutions probed at the distances of our galaxies. The irregulars tend to be relatively nearby. It is possible that AGNs in dwarf irregular galaxies are particularly difficult to identify at greater distances through optical selection since more star-formation related emission can be included in the spectroscopic aperture. Indeed, \citet{dickey2019} obtained Keck data of dwarf galaxies in the SDSS and found some to be classified as AGN with higher resolution data.

\section{Discussion}

\subsection{Comparison to Jiang et al.\ (2011)}

Using {\it HST} observations, \citet{jiang} studied the structures of 147 galaxies hosting low-mass BHs from the sample of \citet{greeneho2007}. The objects in the \citet{greeneho2007} sample were selected as broad-line AGNs in the SDSS with virial BH masses $M_{\rm BH} \lesssim 2 \times 10^6~M_\odot$. The hosts are sub-$L_\star$ galaxies, yet they are more luminous and massive than the dwarfs studied here.   

In some ways, our results are similar to those of \citet{jiang}. For example,
most of their sample is dominated by disk galaxies with small bulge components. Our median $B/T$ ratio of 0.21 $\pm$ 0.03 is comparable to the median $B/T$ ratio of 0.18 reported by \citet{jiang}.
They also found that the vast majority of these bulges were likely to be pseudobulges, rather than classical bulges, which is in agreement with our results. \citet{jiang} report $70\%$ of their modeled (pseudo)bulge components have a S{\'e}rsic index $n < 2$, and we find $\sim 81\%$ of our modeled (pseudo)bulges have a S{\'e}rsic index $n < 2$.

There are also notable differences between our sample of dwarf galaxies and the sample studied by \citet{jiang}. First, we see a very small fraction ($\sim 11\%$) of galaxies with bars. In contrast, \citet{jiang} report a bar in $\sim 39\%$ of their sample. It has been postulated \citep[e.g.,][]{Shlosman} that bars play a role in funneling gas to the centers of galaxies and feeding AGN. Our sample suggests there must be more at play than just bars. This same conclusion was reached by \citet{jiang}; despite their sample having a significant bar fraction, the fraction was still too low to suggest that bars alone feed AGN. Finally, in contrast with \citet{jiang}, we find dwarf irregular galaxies in our sample (\S\ref{sec:irregulars}). 

\subsection{Nature of the Point Sources: Nuclear Star Clusters and/or AGNs?}\label{sec:nsc}

Here we consider the physical origin of the PSFs used in our models. Are the unresolved point sources dominated by the AGNs in the galaxies, or nuclear star clusters (NSCs)? The dwarf galaxies in our sample were selected to have optical signatures of AGNs \citep{reines}, and $\sim 80\%$ of galaxies in this stellar mass range ($M_\star \sim 10^{8.5} - 10^{9.5} M_\odot$) are known to host NSCs (e.g., see Figure 3 in the review by \citealt{neumayer}). Indeed, AGNs and NSCs are known to co-exist in many galaxies \citep{seth2008}.

Both types of objects would appear as point sources in our {\it HST} WFC3 F110W images. The angular resolution is 0\farcs13, corresponding to a physical resolution of $\sim$ 70 pc at the median distance of our sample. Even in the nearest modeled galaxy in our sample (RGG 94, with a redshift of 0.0066), a point source corresponds to a physical size of $\lesssim$ 17 pc. NSCs tend to have radii $\lesssim$ 10-15 pc \citep{gehansc, neumayer}, and therefore an NSC would likely appear as an unresolved source of light for all galaxies in our sample. Of course, continuum emission from AGNs would also be unresolved on these scales. 

The F110W luminosities of the PSFs in our SNAP sample range from $10^{39.6}$ to $10^{42.1}$ erg s$^{-1}$, with a median of $<L_{\rm PSF}> = 10^{41.1}$ erg s$^{-1}$. We estimate the expected NSC luminosity in each galaxy using the scaling between galaxy stellar mass and NSC mass given by \citet{neumayer}:

\begin{equation}\label{eqn:nsc}
  {\rm log}~M_{\rm NSC} = 0.48 \times {\rm log}\left(\frac{M_*}{10^9M_\odot}\right) + 6.51
\end{equation}

\noindent
 Galaxy stellar masses are adopted from the NSA and given in Table 1. We estimate the F110W luminosity given the predicted NSC mass with Starbust99 models for the continuum \citep{leitherer}, assuming an instantaneous burst and a metallicity of 0.008 as appropriate for these dwarfs \citep{reines}. 
 
\begin{figure}
\includegraphics[width=0.5\textwidth]{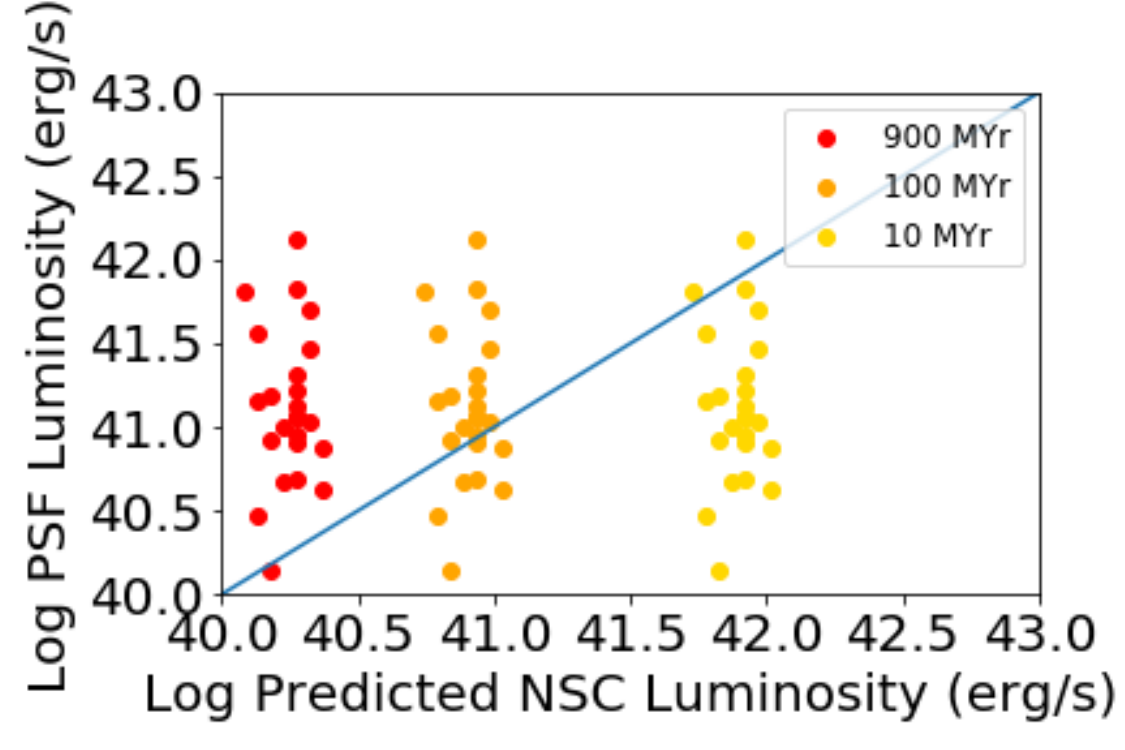}
\caption{Central PSF luminosity for dwarf galaxies in our SNAP sample versus predicted NSC luminosity for various stellar population ages (see \S\ref{sec:nsc}). The blue line shows the one-to-one relation.}
\label{fig:nsclum}
\end{figure}

 We show the measured PSF luminosities in our galaxies versus the predicted luminosities for stellar populations (i.e., NSCs) of three different ages in Figure \ref{fig:nsclum}. The two older ages, 900 Myr and 100 Myr, are typical of NSCs \citep{neumayer}, while 10 Myr would be an abnormally young age for an NSC. However, we include the 10 Myr model because the most luminous observed PSFs are $\sim10-100 \times$ the predicted NSC luminosity unless we assume such a young cluster. This may suggest these simple stellar population models do not adequately represent NSCs, or that there is another contribution to the unresolved source of light (such as the AGN), at least for the most luminous PSFs. It is also possible that the scaling relation between NSC mass and stellar mass may be bimodal (as discussed in \citealt{fornax}), rather than the relation in Equation \ref{eqn:nsc}.

Naively, we would not expect to detect continuum emission from the narrow-line AGNs in our SNAP sample, since the lack of detectable broad-line emission suggests that the nuclei are obscured. However, a nucleus could be unobscured, with the broad-line emission falling below the detection limit in the \citet{reines} study. This scenario is exemplified by RGG 118, in which broad H$\alpha$ emission was not firmly detected in the SDSS spectrum, but was detected in follow up data with higher sensitivity and spectral resolution \citep{baldassare2015}. It is also possible that the obscuring material is clumpy, keeping some sight lines to the nucleus open that could contribute to the continuum we are observing. Additionally, continuum emission via scattered light has been observed from obscured nuclei \citep{zakamska}, which could be contributing to our point sources. There could also be some contribution at 1.1 $\mu$m from very hot dust.

\subsection{Selection Effects}

We note that our sample of AGN-hosting dwarf galaxies is not complete. The dwarf galaxies studied here were selected to have optical spectroscopic indications of an AGN \citep{reines}. There are other ways to identify active BHs in dwarf galaxies - e.g., radio \citep{reines2011,reines2014,reines2020}, X-ray \citep{lemons2015,birchall2020}, AGN variability \citep{baldassare2020}, and the host properties can be quite different depending on the BH selection method \citep[e.g.,][]{reines2020}.

As discussed in \citet{reines}, optical diagnostics are only sensitive to actively accreting BHs in galaxies with relatively low star formation. Even if accreting at the Eddington limit, lower-mass BHs are not very luminous, leading to a selection bias towards dwarf galaxies hosting highly accreting BHs and/or more massive BHs.  Emission lines from an AGN can also be hidden by host galaxy light even without significant star formation \citep{moran2002}. Moreover, low-metallicity AGNs overlap with low-metallicity starbursts in the upper-left region of the BPT diagram \citep[e.g.,][]{groves2006,yang2017}. This will lead to a bias against actively star-forming dwarf galaxies, despite possibly hosting a massive BH \citep{reines2011,riffel2020,birchall2020,baldassare2020}. The 3\arcsec- diameter SDSS fibers are three times the median angular effective diameter of the (pseudo)bulge components of our galaxies (median $2r_{e,{\rm bulge}}$ = 1\arcsec); therefore only galaxies which are AGN dominant and have bright, well-defined centers will be selected via SDSS spectroscopy \citep{reines}. 

In addition, many dwarf galaxies are simply too faint to be targeted for spectroscopy in the SDSS and even if they are, there is no guarantee the fiber placement coincides with a potential AGN. Instead, the spectroscopic fiber may be centered on a bright star forming region, and/or the BH may not reside in the nucleus \citep{reines2020}.
Therefore, while the \citet{reines} sample is likely highly incomplete, the galaxies studied here are representative of that optically-selected sample of AGNs in dwarf galaxies (Figure \ref{fig:sampleproperties}).

\subsection{Environmental Effects on Galaxy Morphologies}
It has been observed that the environment around a galaxy plays a role in the morphology of the galaxy. Galaxies which are in less dense environments tend to be diskier, while galaxies in a more tightly packed environment tend to be rounder and more bulge-like (see, e.g. \citealt{rong2020}).

For 28 of the dwarf galaxies in this work, the distance to the nearest massive host galaxy has been measured using the 2MASS Extended Source Catalog and taking redshifts from SDSS spectroscopy and several other sources \citep{gehaquench}. The remaining galaxies fall within one degree of the edge of the SDSS footprint, and so the environment was not analyzed. The nearest massive host galaxy was chosen to be the nearest galaxy with $M_{K_s} < -23$, and dwarfs were considered to be ``isolated" by \citet{gehaquench} if the nearest massive host is more than 1.5 Mpc away. This is a conservative estimate of whether a galaxy is isolated or not; many galaxies less than 1.5 Mpc away from a more massive galaxy are still isolated, as they are often separated by many virial radii.

\begin{figure}
\includegraphics[width=0.5\textwidth]{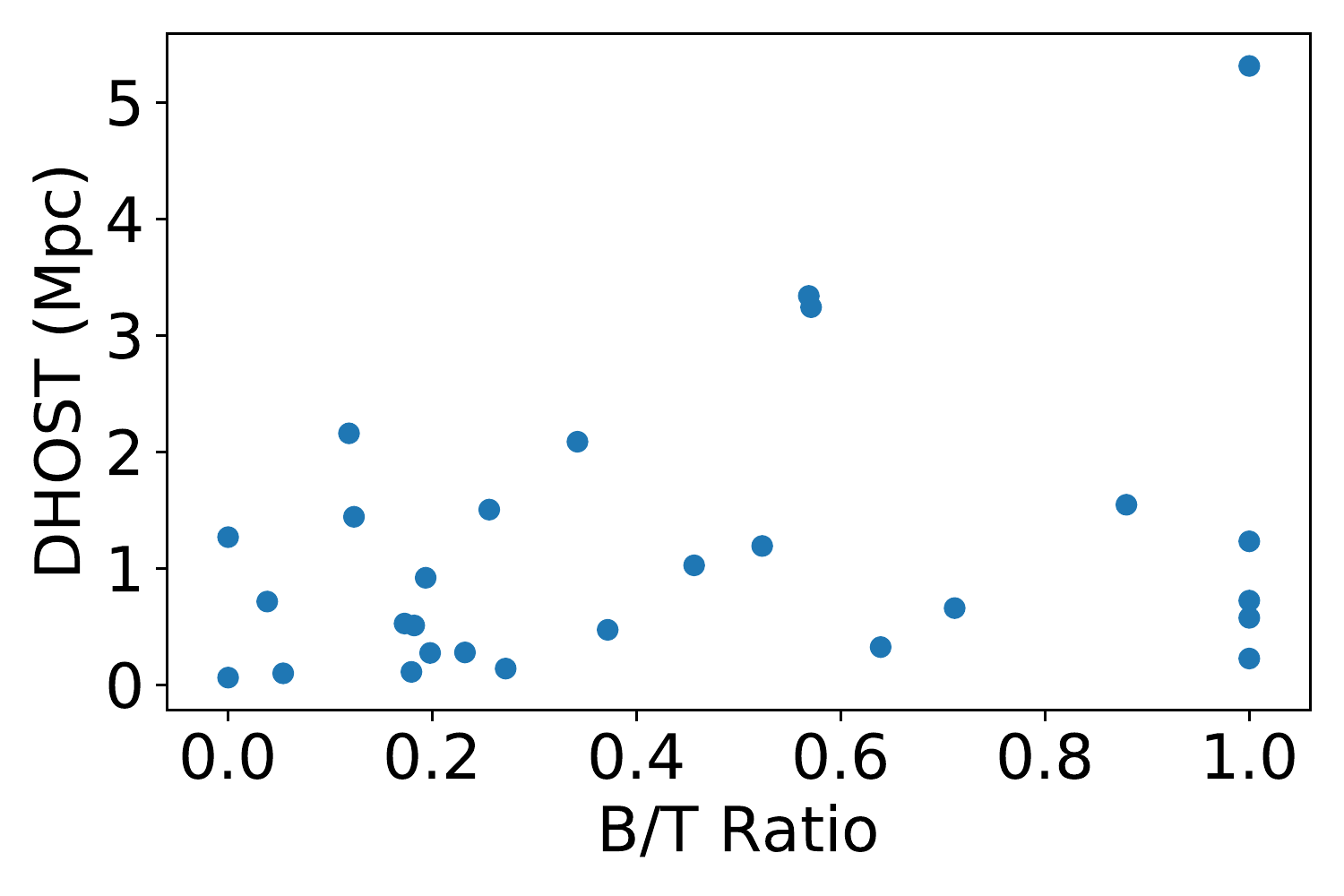}
\caption{Distance from the dwarf galaxy to the nearest massive host versus the measured bulge to total light ratio.}
\label{fig:dhost}
\end{figure}

We show the B/T ratio plotted against the distance to the nearest host (DHOST) for the galaxies with environment data in Figure \ref{fig:dhost}. We find that the majority of dwarfs in our sample do not meet the criterion to be considered isolated, but we do not find a correlation between whether the bulge or disk dominates and the isolation of the galaxy. For our sample, it may be difficult to disentangle the effects of hosting an AGN from the effects of the environment.

\section{Conclusions}

We have presented a study of 41 dwarf galaxies hosting optically-selected massive BHs \citep{reines} using {\it HST} near-infrared observations. In this first paper, we examine the morphologies and structural components of the host galaxies using the galaxy image fitting software GALFIT \citep{peng}. We will compare these AGN-hosting dwarf galaxies to the general population of dwarf galaxies in a forthcoming paper.
The main results of this work are summarized below:

\begin{enumerate}
  \item The majority of the dwarf galaxies in our sample (85\%) have regular morphologies, although there is a non-negligible fraction (15\%) of irregular/disturbed galaxies in our sample (see Figure \ref{fig:irrimgs}).
  \item We perform 2D bulge-disk decompositions for the regular galaxies and find that the majority are disk-dominated with small pseudobulges. The median bulge-to-total light ratio is $<B/T>=0.21$.
  \item Our sample also includes three dwarf disk galaxies without detectable bulges and six pure bulge/elliptical galaxies.
  \item The best-fit models for the regular dwarf galaxies include a central point source of light. The point sources are consistent with originating from nuclear star clusters and/or AGNs. 
  \item Of the irregular/disturbed galaxies, three appear to be Magellanic-type dwarf irregulars and two exhibit obvious tidal features indicative of interactions/mergers.
\end{enumerate}

We have shown that optically-selected BH-hosting dwarf galaxies exhibit a variety of morphologies and structures. This has important implications for constraining the BH occupation fraction at low mass, which is a key diagnostic for discriminating between BH seeding mechanisms \citep[e.g.,][]{volonteri,greenetal2019}. While there have been valiant efforts to constrain the BH occupation fraction in low mass galaxies \citep{miller2015}, studies that only focus on a particular type of dwarf galaxy (e.g., early-types) may well miss the bulk of the population.

\acknowledgements

We thank the anonymous referee for their helpful comments and questions during the review process. SJK acknowledges support for this project provided by a fellowship from the Montana Space Grant Consortium.
Support for Program number HST-GO-14251.004A was provided by NASA through a grant from the Space Telescope Science Institute, which is operated by the Association of Universities for Research in Astronomy, Incorporated, under NASA contract NAS5-26555.
AER gratefully acknowledges support for this paper provided by NASA through EPSCoR grant number 80NSSC20M0231.

\clearpage
\appendix

\section{HST images and GALFIT models}

\begin{figure*}[!h]
$\begin{array}{c}
{\includegraphics[width=\textwidth]{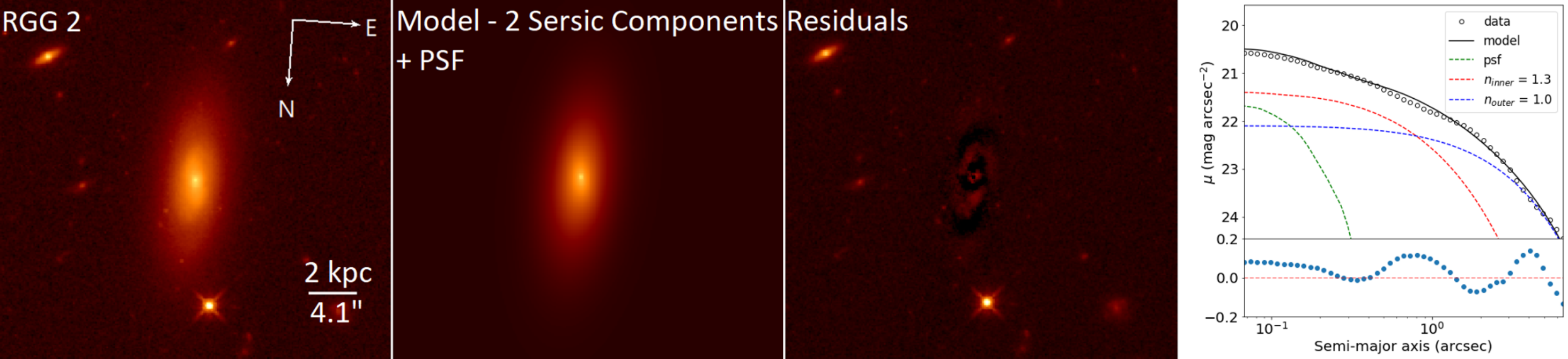}} \\
{\includegraphics[width=\textwidth]{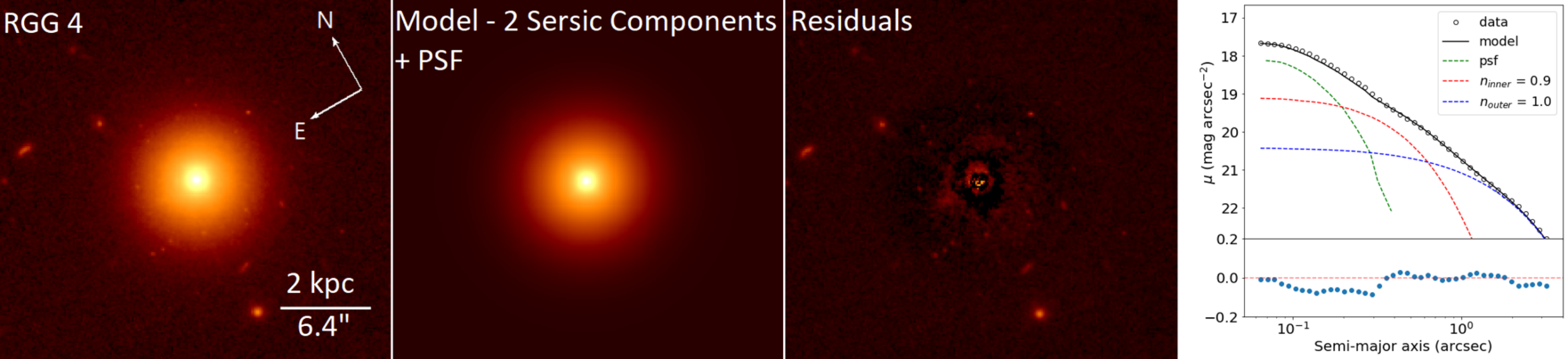}} \\
{\includegraphics[width=\textwidth]{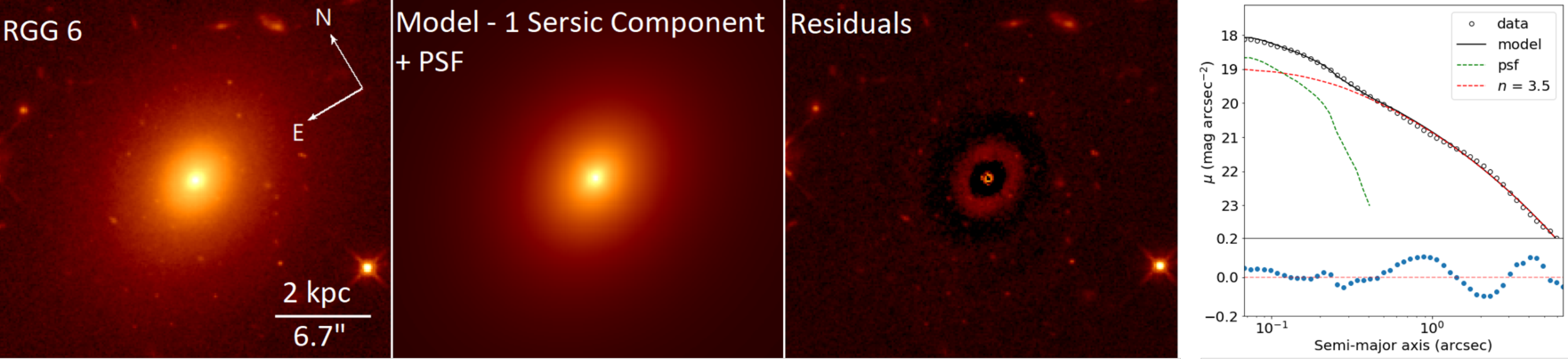}} \\
{\includegraphics[width=\textwidth]{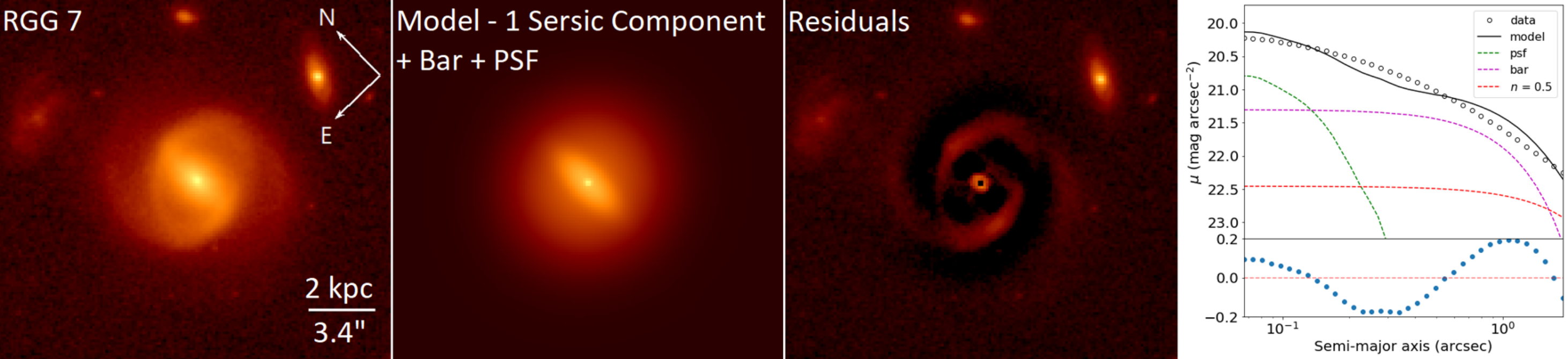}} \\
{\includegraphics[width=\textwidth]{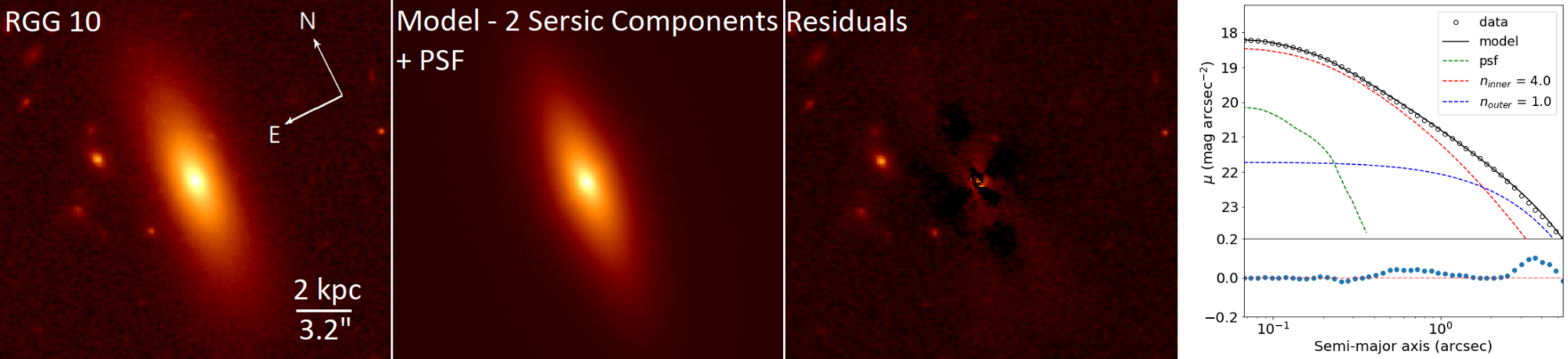}}
\end{array}$
\caption{Left three panels: {\it HST} image, GALFIT model, and the residuals after subtracting the model from the image. Images are shown on a stretched log scale to show faint details in the residuals. Right: Surface brightness profiles. The data are shown as black circles and the model is shown as a black line. The individual model components are shown as colored dashed lines. The residuals are shown in the bottom panel.}
\label{fig:appendix1}
\end{figure*}

\newpage

\begin{figure*}[!h]
$\begin{array}{c}
{\includegraphics[width=\textwidth]{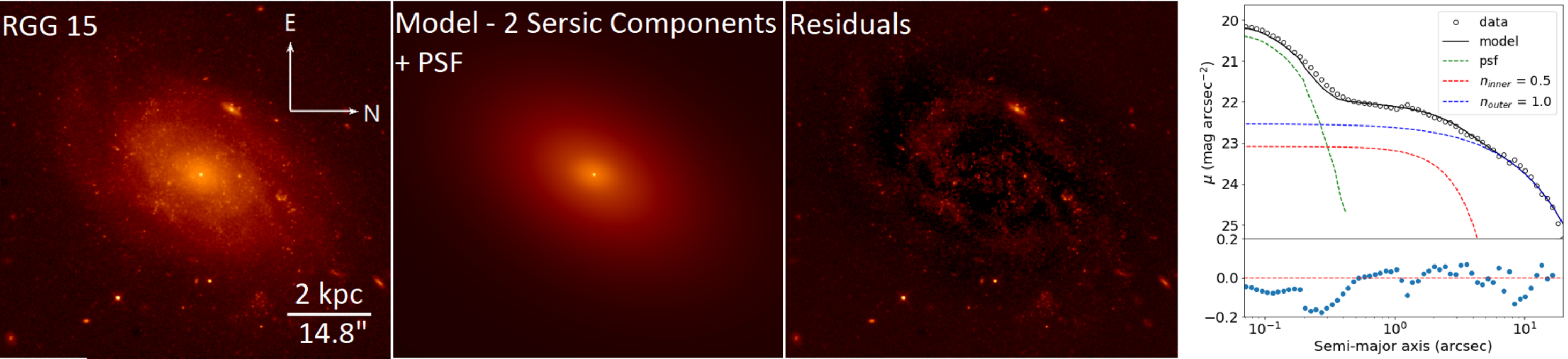}} \\
{\includegraphics[width=\textwidth]{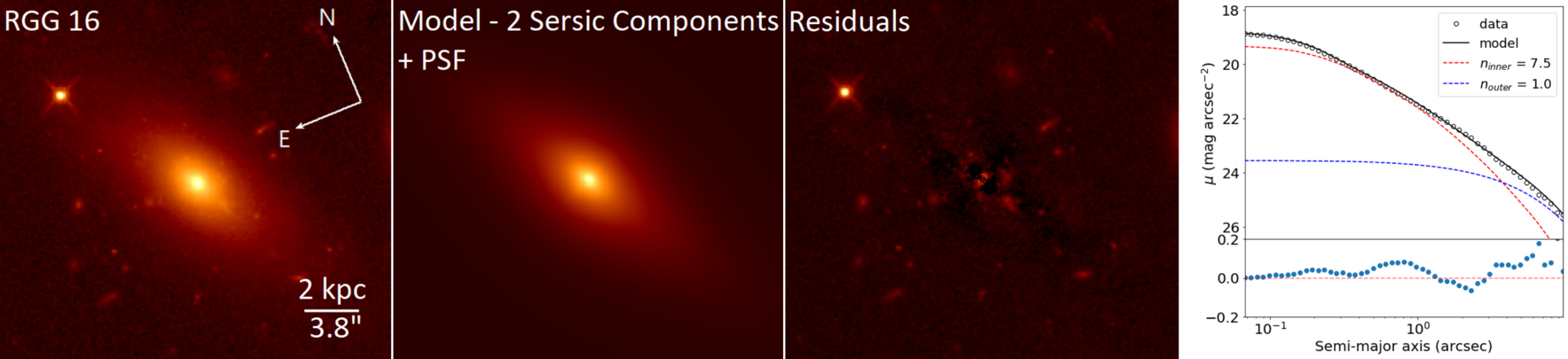}} \\
{\includegraphics[width=\textwidth]{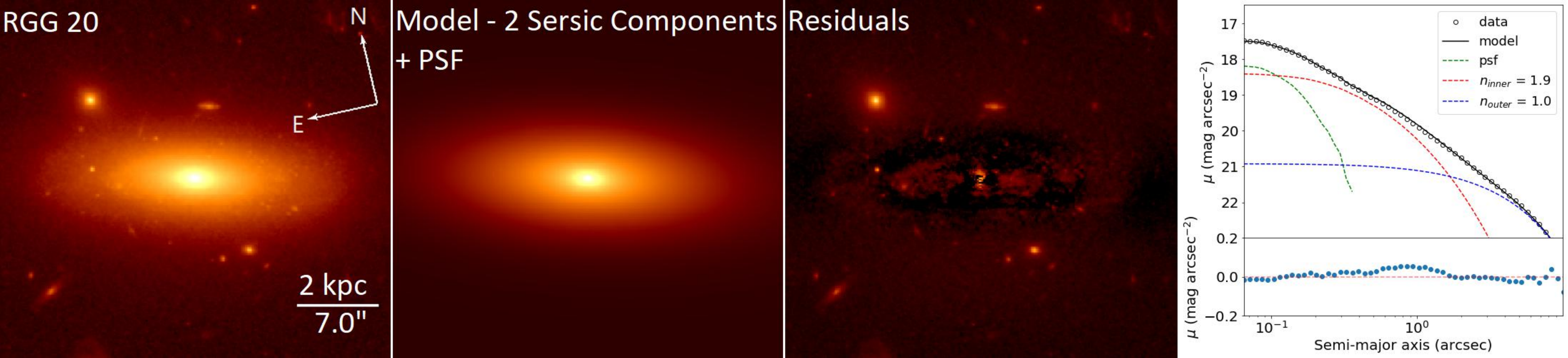}} \\
{\includegraphics[width=\textwidth]{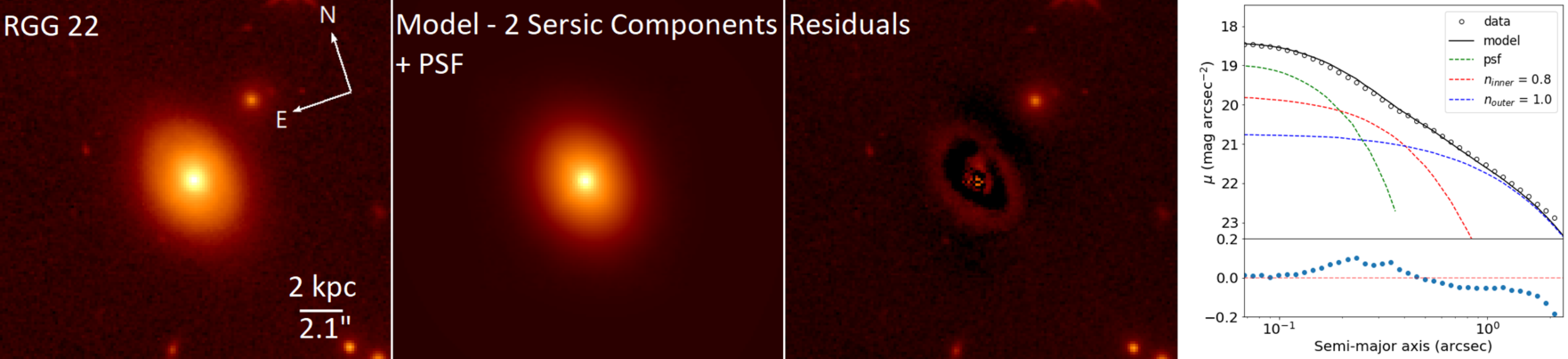}} \\
{\includegraphics[width=\textwidth]{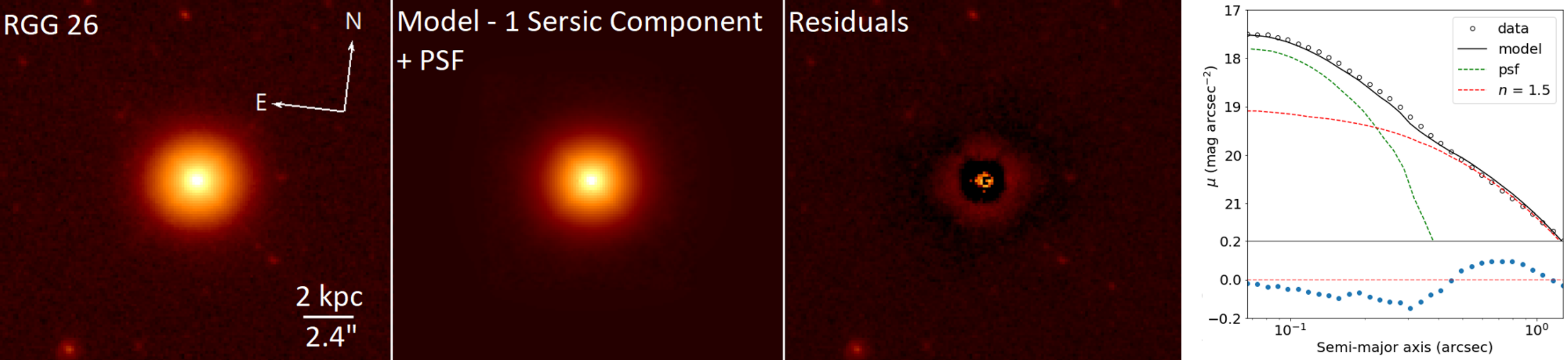}}
\end{array}$
\caption{Left three panels: {\it HST} image, GALFIT model, and the residuals after subtracting the model from the image. Images are shown on a stretched log scale to show faint details in the residuals. Right: Surface brightness profiles. The data are shown as black circles and the model is shown as a black line. The individual model components are shown as colored dashed lines. The residuals are shown in the bottom panel.}
\label{fig:appendix2}
\end{figure*}

\begin{figure*}[!h]
$\begin{array}{c}
{\includegraphics[width=\textwidth]{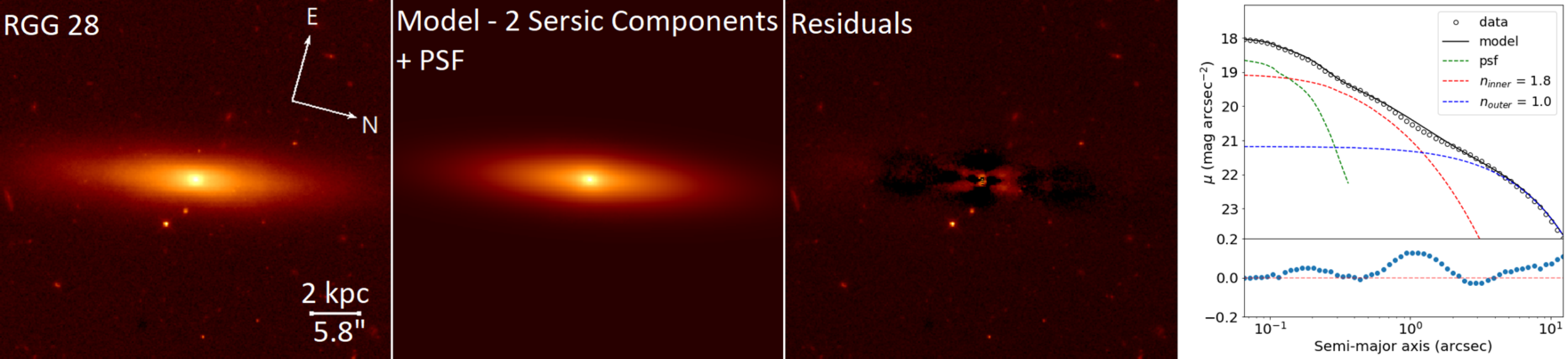}} \\
{\includegraphics[width=\textwidth]{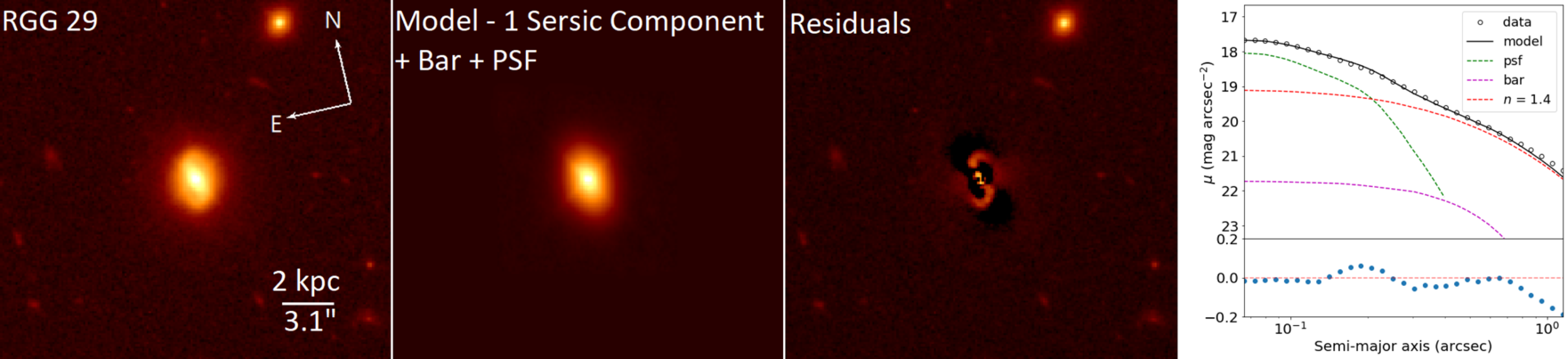}} \\
{\includegraphics[width=\textwidth]{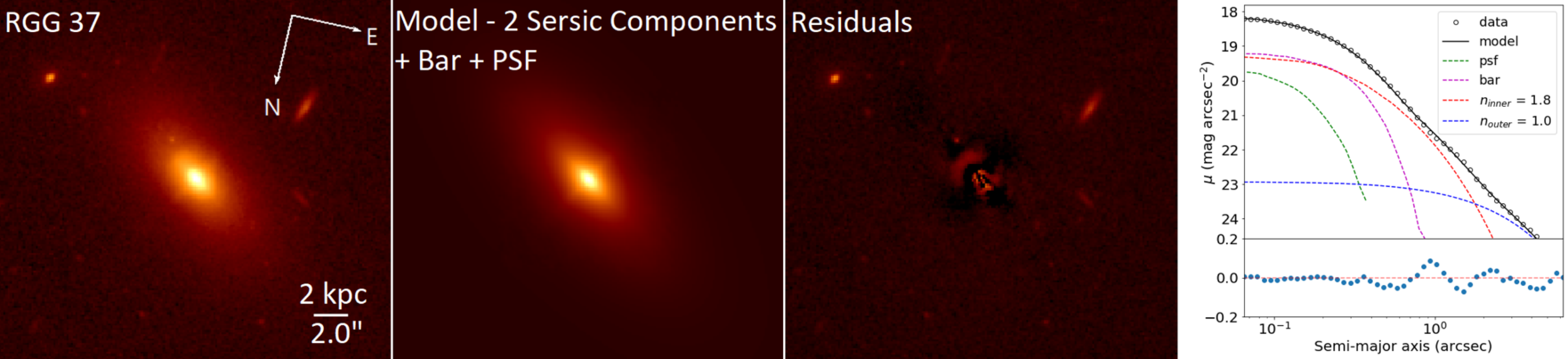}} \\
{\includegraphics[width=\textwidth]{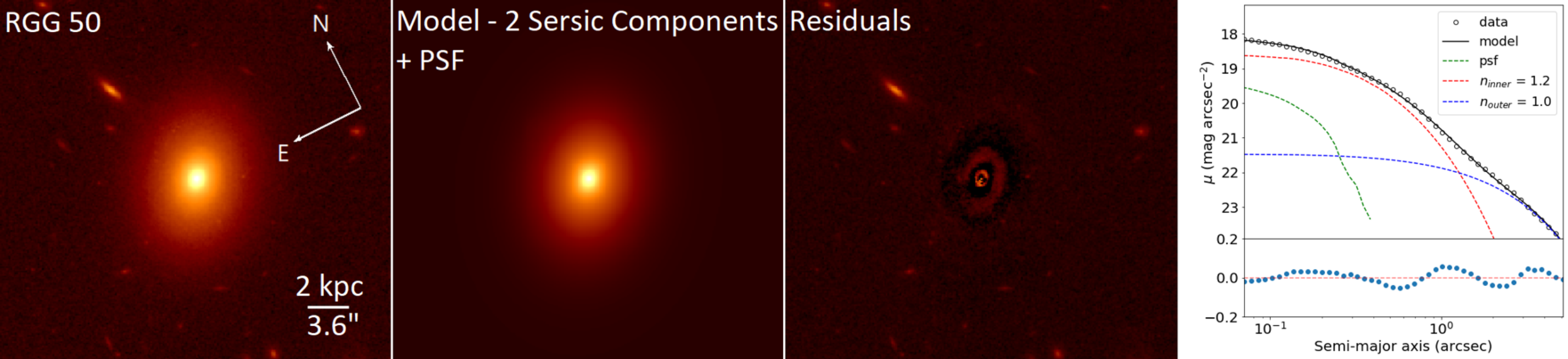}} \\
{\includegraphics[width=\textwidth]{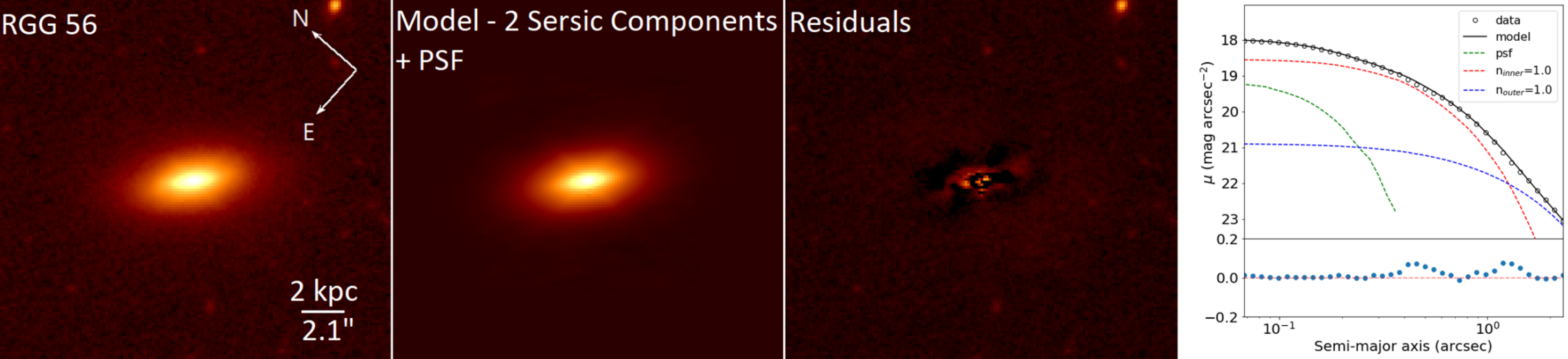}}
\end{array}$
\caption{Left three panels: {\it HST} image, GALFIT model, and the residuals after subtracting the model from the image. Images are shown on a stretched log scale to show faint details in the residuals. Right: Surface brightness profiles. The data are shown as black circles and the model is shown as a black line. The individual model components are shown as colored dashed lines. The residuals are shown in the bottom panel.}
\label{fig:appendix3}
\end{figure*}

\begin{figure*}[!h]
$\begin{array}{c}
{\includegraphics[width=\textwidth]{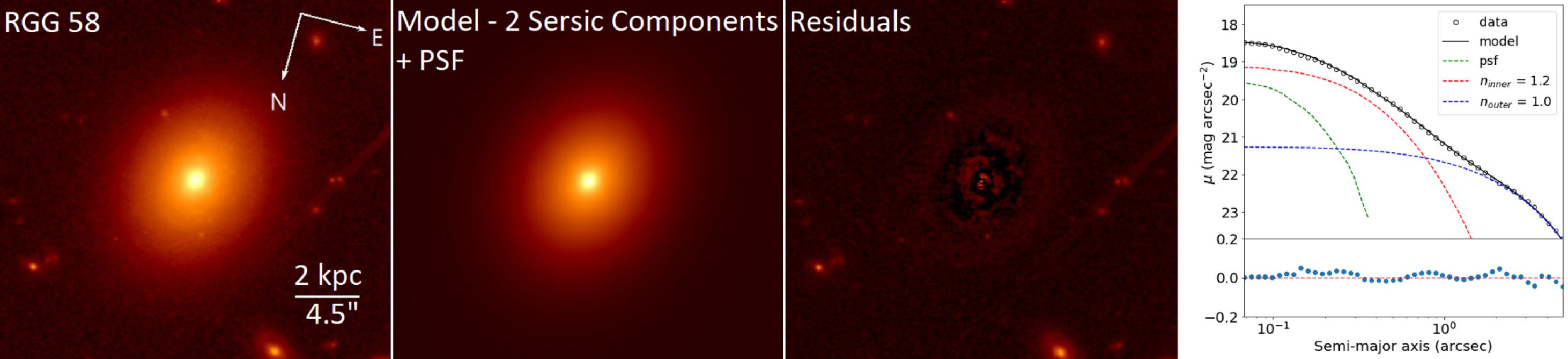}} \\
{\includegraphics[width=\textwidth]{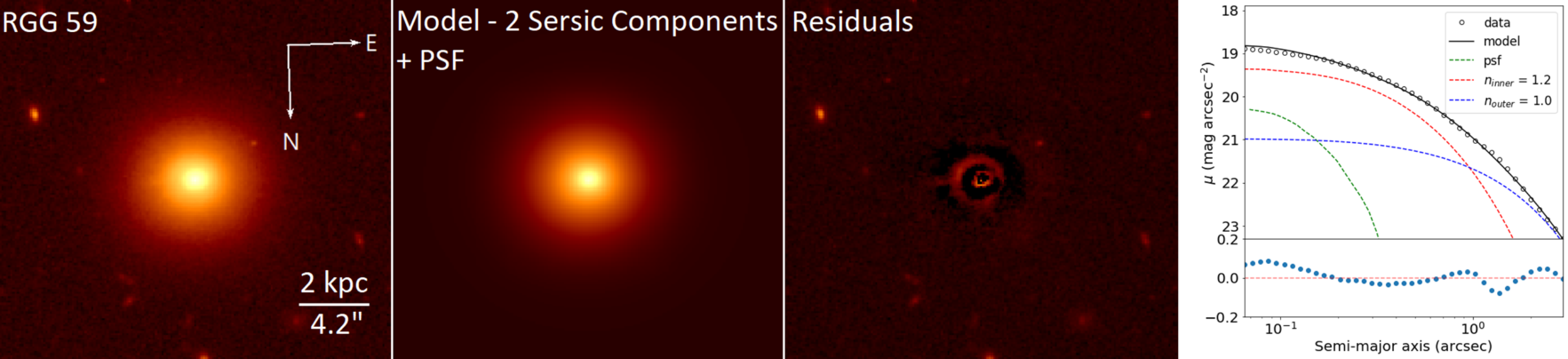}} \\
{\includegraphics[width=\textwidth]{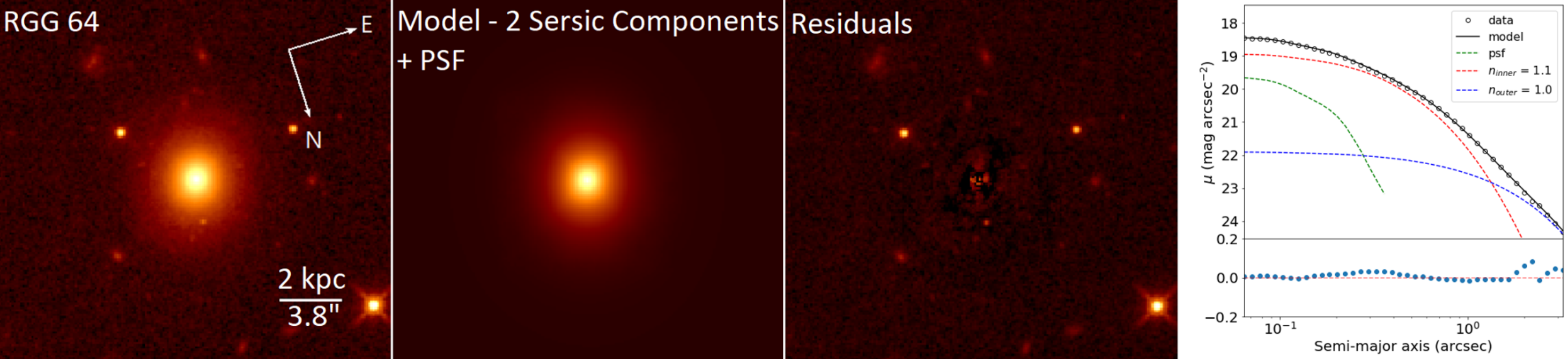}} \\
{\includegraphics[width=\textwidth]{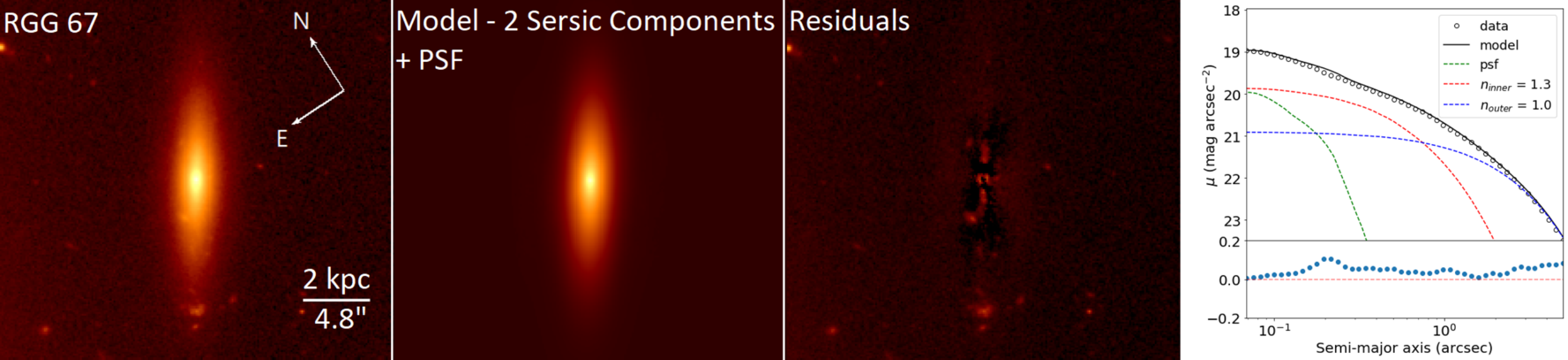}} \\
{\includegraphics[width=\textwidth]{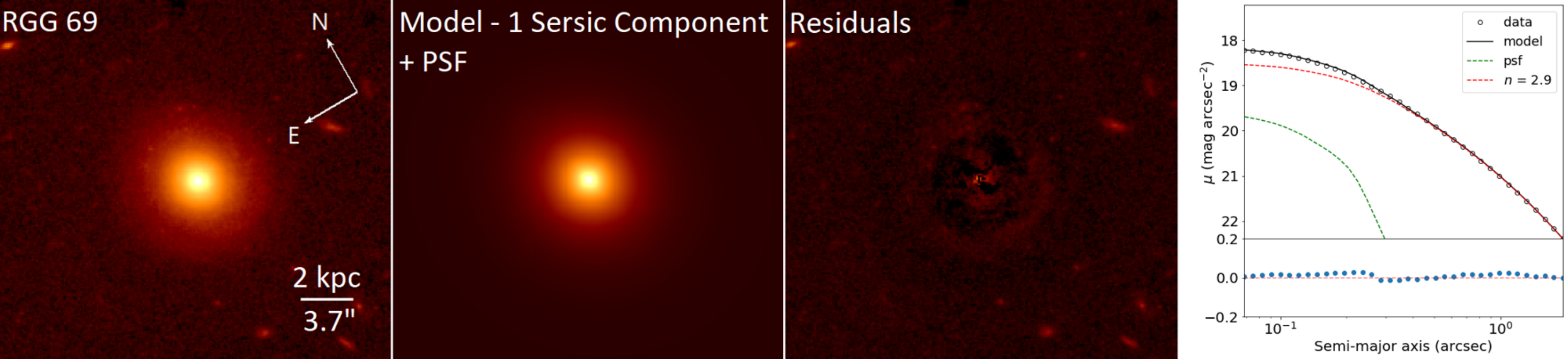}}
\end{array}$
\caption{Left three panels: {\it HST} image, GALFIT model, and the residuals after subtracting the model from the image. Images are shown on a stretched log scale to show faint details in the residuals. Right: Surface brightness profiles. The data are shown as black circles and the model is shown as a black line. The individual model components are shown as colored dashed lines. The residuals are shown in the bottom panel.}
\label{fig:appendix4}
\end{figure*}

\begin{figure*}[!h]
$\begin{array}{c}
{\includegraphics[width=\textwidth]{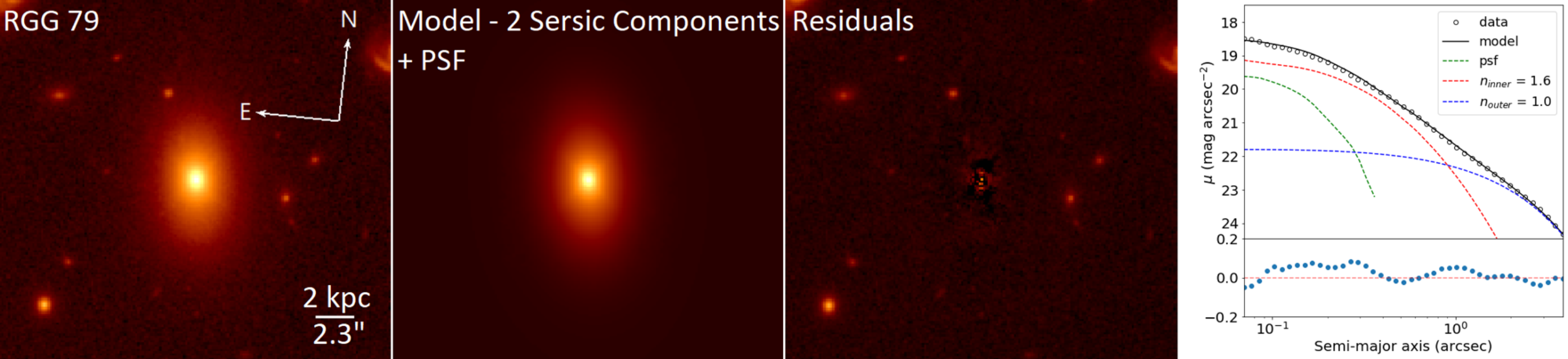}} \\
{\includegraphics[width=\textwidth]{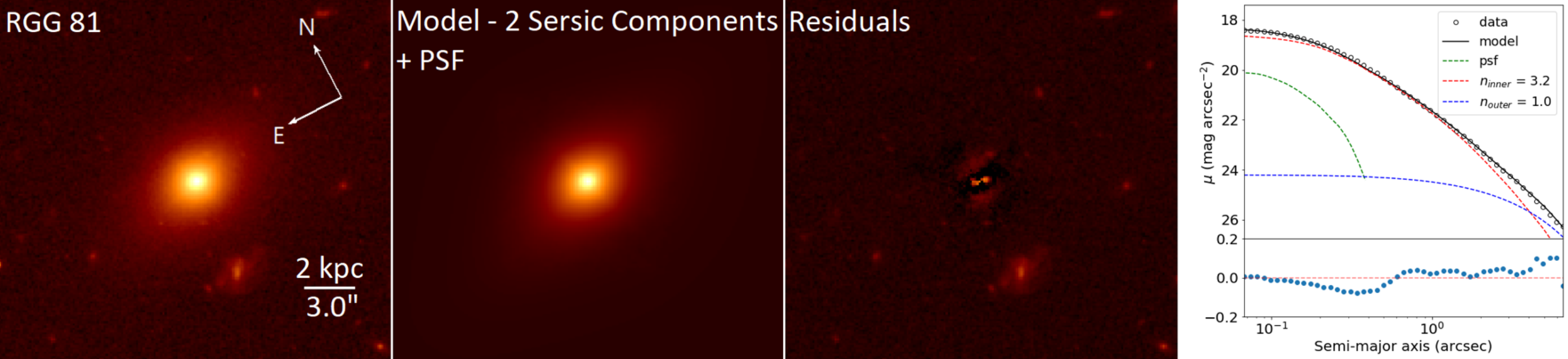}} \\
{\includegraphics[width=\textwidth]{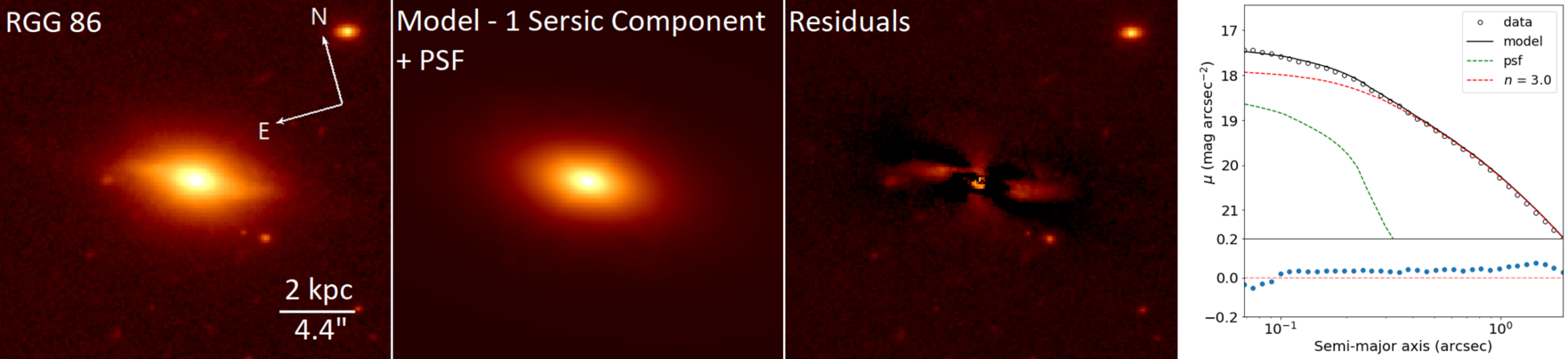}} \\
{\includegraphics[width=\textwidth]{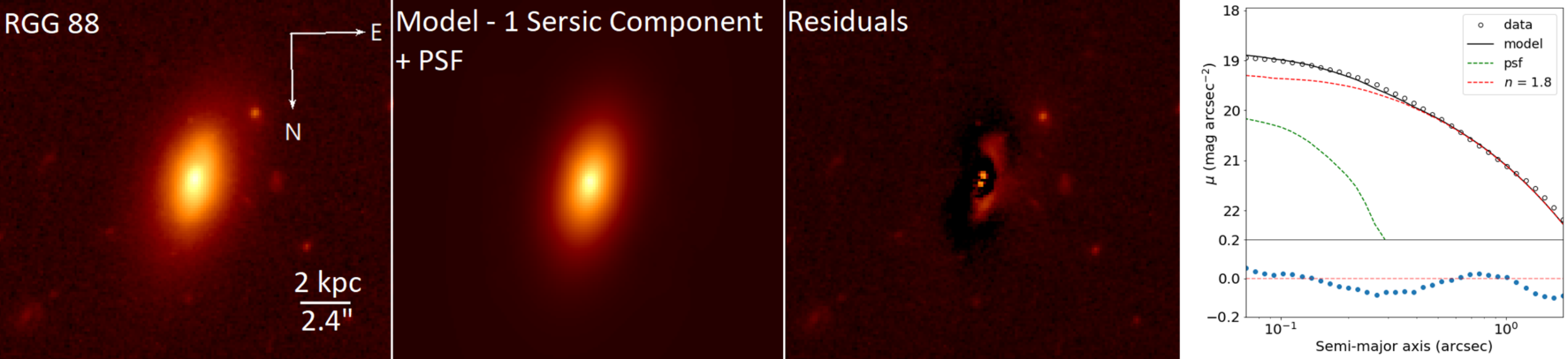}} \\
{\includegraphics[width=\textwidth]{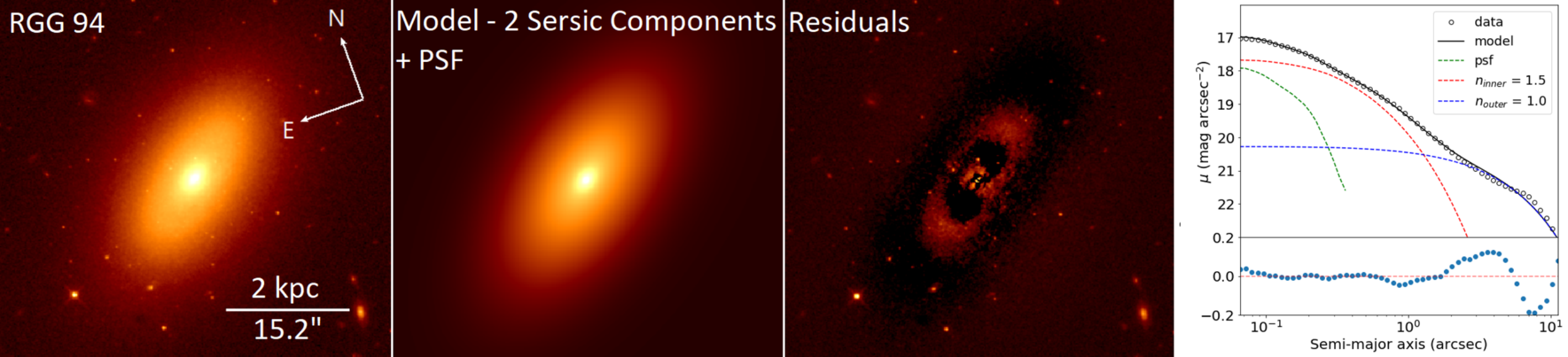}}
\end{array}$
\caption{Left three panels: {\it HST} image, GALFIT model, and the residuals after subtracting the model from the image. Images are shown on a stretched log scale to show faint details in the residuals. Right: Surface brightness profiles. The data are shown as black circles and the model is shown as a black line. The individual model components are shown as colored dashed lines. The residuals are shown in the bottom panel.}
\label{fig:appendix5}
\end{figure*}

\clearpage

\LongTables
\begin{deluxetable*}{ccccccc}[!h]
\tabletypesize{\footnotesize}
\tablewidth{0pt}
\tablecaption{Fitting Results for Galaxies With Regular Morphologies \label{table:secureresults}}
\tablehead{
\colhead{RGG ID} & \colhead{Component} & \colhead{\textit{m}$_{F110W}$} & \colhead{n} & \colhead{$R_{e}$ (kpc)} & \colhead{q} & \colhead{Additional Components} \\
\colhead{(1)} & \colhead{(2)} & \colhead{(3)} & \colhead{(4)} & \colhead{(5)} & \colhead{(6)} & \colhead{(7)}}
\startdata
& PSF & 21.86 $\pm$ 0.02 & - & - & - \\
RGG 1$^a$ & Inner S{\'e}rsic & 19.52 $\pm$ 0.23 & 0.32 $\pm$ 0.09 & 0.71 $\pm$ 0.03 & 0.51 \\
& Outer S{\'e}rsic & 17.41 $\pm$ 0.06 & 0.83 $\pm$ 0.13 & 1.61 $\pm$ 0.03 & 0.72\\
\\
& PSF & 23.9 $\pm$ 0.01 & - & - & - \\
RGG 2 & Inner S{\'e}rsic & 20.3 $\pm$ 0.15 & 1.31 $\pm$ 0.01 & 0.73 $\pm$ 0.01 & 0.44 & - \\
& Outer S{\'e}rsic & 18.78 $\pm$ 0.10 & 1.00 & 2.16 $\pm$ 0.01 & 0.41 & - \\
\\
& PSF & 20.41 $\pm$ 0.01 & - & - & - \\
 RGG 4 & Inner S{\'e}rsic & 19.22 $\pm$ 0.06 & 0.87 $\pm$ 0.01 & 0.13 $\pm$ 0.01 & 0.95\\
 & Outer S{\'e}rsic & 17.67 $\pm$ 0.03 & 1.00 & 0.68 $\pm$ 0.01 & 0.99\\
 \\

& PSF & 20.99 $\pm$ 0.05 & - & - & - \\
RGG 6 & S{\'e}rsic & 17.05 $\pm$ 0.10 & 3.52 $\pm$ 0.06 & 1.32 $\pm$ 0.04 & 0.81 \\
\\

& PSF & 23.07 $\pm$ 0.01 & - & - & - \\
RGG 7 & S{\'e}rsic & 18.93 $\pm$ 0.02 & 0.52 $\pm$ 0.01 & 1.41 $\pm$ 0.01 & 0.96 & Bar ($\textit{m}_{F110W} = 20.09$) \\
\\

& PSF & 21.86 $\pm$ 0.02 & - & - & - \\
RGG 9$^a$ & S{\'e}rsic & 17.01 $\pm$ 0.04 & 2.30 $\pm$ 0.11 & 1.21 $\pm$ 0.26 & 0.86 \\
\\

& PSF & 22.39 $\pm$ 0.05 & - & - & - \\
RGG 10 & Inner S{\'e}rsic & 18.51 $\pm$ 0.01 & 3.97 $\pm$ 0.78 & 0.72 $\pm$ 0.15 & 0.48 \\
& Outer S{\'e}rsic & 19.13 $\pm$ 0.05 & 1.00 & 2.21 $\pm$ 0.09 & 0.29\\
\\

& PSF & 19.61 $\pm$ 0.03 & - & - & - \\
RGG 11$^a$ & Inner S{\'e}rsic & 18.40 $\pm$ 0.23 & 2.40 $\pm$ 0.18 & 0.13 $\pm$ 0.02 & 0.96 \\
& Outer S{\'e}rsic & 16.22 $\pm$ 0.18 & 1.69 $\pm$ 0.09 & 2.57 $\pm$ 0.30 & 0.78 \\
\\

& PSF & 22.54 $\pm$ 0.02 & - & - & - \\
RGG 15 & Inner S{\'e}rsic & 19.8 $\pm$ 0.17 & 0.48 $\pm$ 0.01 & 0.33 $\pm$ 0.01 & 0.70 \\
& Outer S{\'e}rsic & 16.3 $\pm$ 0.01 & 1.00 & 1.96 $\pm$ 0.01 & 0.64 \\
\\

& PSF & 22.17 $\pm$ 0.01 & - & - & - \\
RGG 16 & Inner S{\'e}rsic & 18.52 $\pm$ 0.15 & 3.32 $\pm$ 0.22 & 0.98 $\pm$ 0.08 & 0.68\\
& Outer S{\'e}rsic & 19.50 $\pm$ 0.07 & 1.00 & 3.88 $\pm$ 0.04 & 0.30 \\
\\

& PSF & 20.45 $\pm$ 0.01 & - & - & - \\
RGG 20 & Inner S{\'e}rsic & 17.94 $\pm$ 0.15 & 1.86 $\pm$ 0.01 & 0.33 $\pm$ 0.02 & 0.46\\
& Outer S{\'e}rsic & 16.87 $\pm$ 0.07 & 1.00 & 1.93 $\pm$ 0.02 & 0.36 \\
\\

& PSF & 21.18 $\pm$ 0.01 & - & - & - \\
RGG 22 & Inner S{\'e}rsic & 20.55 $\pm$ 0.09 & 0.83 $\pm$ 0.05 & 0.28 $\pm$ 0.01 & 0.83 \\
& Outer S{\'e}rsic & 18.9 $\pm$ 0.04 & 1.00 & 1.45 $\pm$ 0.03 & 0.78\\
\\

& PSF & 20.08 $\pm$ 0.01 & - & - & - \\
RGG 26 & S{\'e}rsic & 18.54 $\pm$ 0.01 & 1.54 $\pm$ 0.09 & 0.68 $\pm$ 0.01 & 0.91\\
\\

& PSF & 20.86 $\pm$ 0.02 & - & - & - & - \\
RGG 28 & Inner S{\'e}rsic & 18.54 $\pm$ 0.03 & 1.77 $\pm$ 0.02 & 0.38 $\pm$ 0.01 & 0.56\\
& Outer S{\'e}rsic & 17.24 $\pm$ 0.02 & 1.00 & 2.83 $\pm$ 0.02 & 0.21\\
\\

& PSF & 20.32 $\pm$ 0.01 & - & - & - \\
RGG 29 & S{\'e}rsic & 18.97 $\pm$ 0.01 & 1.40 $\pm$ 0.06 & 0.47 $\pm$ 0.01 & 0.61 & Bar ($\textit{m}_{F110W} = 22.26$)\\
\\

& PSF & 18.45 $\pm$ 0.03 & - & - & - \\
RGG 32$^a$ & Inner S{\'e}rsic & 17.77 $\pm$ 0.25 & 1.62 $\pm$ 0.20 & 0.29 $\pm$ 0.02 & 0.90 \\
& Outer S{\'e}rsic & 16.07 $\pm$ 0.10 & 0.74 $\pm$ 0.03 & 2.03 $\pm$ 0.01 & 0.95 \\
\\

& PSF & 21.96 $\pm$ 0.08 & - & - & - \\
RGG 37 & Inner S{\'e}rsic & 19.50 $\pm$ 0.01 & 1.79 $\pm$ 0.16 & 0.66 $\pm$ 0.01 & 0.53 & Bar ($\textit{m}_{F110W} = 20.66$) \\
& Outer S{\'e}rsic & 19.60 $\pm$ 0.28 & 1.00 & 4.27 $\pm$ 0.15 & 0.47\\
\\

& PSF & 21.55 $\pm$ 0.01 & - & - & - \\
RGG 48$^a$ & Inner S{\'e}rsic & 19.75 $\pm$ 0.07 & 0.61 $\pm$ 0.08 & 0.29 $\pm$ 0.06 & 0.42 \\
& Outer S{\'e}rsic & 16.64 $\pm$ 0.06 & 0.29 $\pm$ 0.01 & 2.12 $\pm$ 0.30 & 0.48 \\
\\

& PSF & 21.70 $\pm$ 0.10 & - & - & - \\
RGG 50 & Inner S{\'e}rsic & 18.66 $\pm$ 0.21 & 1.24 $\pm$ 0.03 & 0.30 $\pm$ 0.01 & 0.77 \\
& Outer S{\'e}rsic & 18.09 $\pm$ 0.17 & 1.00 & 1.86 $\pm$ 0.01 & 0.75\\
\\

& PSF & 21.42 $\pm$ 0.01 & - & - & - \\
RGG 56 & Inner S{\'e}rsic & 19.19 $\pm$ 0.01 & 1.04 $\pm$ 0.06 & 0.48 $\pm$ 0.01 & 0.26 \\
& Outer S{\'e}rsic & 19.49 $\pm$ 0.02 & 1.00 & 1.48 $\pm$ 0.08 & 0.44\\
\\

& PSF & 21.77 $\pm$ 0.01 & - & - & - \\
RGG 58 & Inner S{\'e}rsic & 19.45 $\pm$ 0.07 & 1.19 $\pm$ 0.01 & 0.19 $\pm$ 0.01 & 0.85 \\
& Outer S{\'e}rsic & 17.82 $\pm$ 0.04 & 1.00 & 1.49 $\pm$ 0.01 & 0.81\\
\\

& PSF & 22.53 $\pm$ 0.01 & - & - & - \\
RGG 59 & Inner S{\'e}rsic & 19.15 $\pm$ 0.15 & 1.19 $\pm$ 0.02 & 0.30 $\pm$ 0.01 & 0.83\\
& Outer S{\'e}rsic & 18.44 $\pm$ 0.07 & 1.00 & 0.96 $\pm$ 0.02 & 0.92 \\
\\

& PSF & 21.87 $\pm$ 0.02 & - & - & - \\
RGG 64 & Inner S{\'e}rsic & 19.05 $\pm$ 0.14 & 1.13 $\pm$ 0.01 & 0.26 $\pm$ 0.01 & 0.87 \\
& Outer S{\'e}rsic & 19.36 $\pm$ 0.20 & 1.00 & 1.13 $\pm$ 0.04 & 0.82\\
\\

& PSF & 22.24 $\pm$ 0.01 & - & - & - \\
RGG 67 & Inner S{\'e}rsic & 19.74 $\pm$ 0.15 & 1.29 $\pm$ 0.06 & 0.36 $\pm$ 0.01 & 0.66\\
& Outer S{\'e}rsic & 18.58 $\pm$ 0.07 & 1.00 & 1.38 $\pm$ 0.01 & 0.26 \\
\\

& PSF & 21.85 $\pm$ 0.01 & - & - & - \\
RGG 69 & S{\'e}rsic & 17.89 $\pm$ 0.15 & 2.88 $\pm$ 0.13 & 0.69 $\pm$ 0.01 & 0.90\\
\\

& PSF & 21.84 $\pm$ 0.02 & - & - & - \\
RGG 79 & Inner S{\'e}rsic & 19.78 $\pm$ 0.18 & 1.61 $\pm$ 0.01 & 0.35 $\pm$ 0.01 & 0.66 \\
& Outer S{\'e}rsic & 19.23 $\pm$ 0.09 & 1.00 & 2.17 $\pm$ 0.04 & 0.60 \\
\\

& PSF & 22.44 $\pm$ 0.13 & - & - & - \\
RGG 81 & Inner S{\'e}rsic & 18.74 $\pm$ 0.13 & 3.17 $\pm$ 0.14 & 0.49 $\pm$ 0.01 & 0.87 \\
& Outer S{\'e}rsic & 20.90 $\pm$ 0.01 & 1.00 & 2.99 $\pm$ 0.05 & 0.40\\
\\

& PSF & 20.79 $\pm$ 0.22 & - & - & - \\
RGG 86 & S{\'e}rsic & 17.68 $\pm$ 0.10 & 2.97 $\pm$ 0.15 & 0.56 $\pm$ 0.01 & 0.49 \\
\\

& PSF & 22.36 $\pm$ 0.22 & - & - & - \\
RGG 88 & S{\'e}rsic & 18.70 $\pm$ 0.10 & 1.79 $\pm$ 0.05 & 0.98 $\pm$ 0.01 & 0.87 \\
\\

& PSF & 21.72 $\pm$ 0.01 & - & - & - \\
RGG 89 & Inner S{\'e}rsic & 19.49 $\pm$ 0.02 & 1.39 $\pm$ 0.03 & 0.31 $\pm$ 0.01 & 0.88 \\
& Outer S{\'e}rsic & 19.30 $\pm$ 0.02 & 1.00 & 1.42 $\pm$ 0.04 & 0.86\\
\\

& PSF & 20.23 $\pm$ 0.01 & - & - & - \\
RGG 94 & Inner S{\'e}rsic & 17.47 $\pm$ 0.02 & 1.45 $\pm$ 0.02 & 0.09 $\pm$ 0.01 & 0.68 \\
& Outer S{\'e}rsic & 15.71 $\pm$ 0.02 & 1.00 & 0.94 $\pm$ 0.01 & 0.54\\
\\

& PSF & 24.26 $\pm$ 0.03 & - & - & - \\
RGG 118$^b$ & Inner S{\'e}rsic & 20.55 $\pm$ 0.09 & 0.80 $\pm$ 0.10 & 1.57 $\pm$ 0.22 & 0.45 \\
& Outer S{\'e}rsic & 18.66 $\pm$ 0.14 & 1.00 & 6.51 $\pm$ 1.72 & 0.69\\
\\

& PSF & 18.94 $\pm$ 0.07 & - & - & - \\
RGG 119$^a$ & Inner S{\'e}rsic & 19.36 $\pm$ 0.21 & 2.55 $\pm$ 0.47 & 0.17 $\pm$ 0.01 & 0.46 \\
& Outer S{\'e}rsic & 17.23 $\pm$ 0.05 & 0.91 $\pm$ 0.06 & 1.02 $\pm$ 0.01 & 0.78 \\
\\

& PSF & 20.00 $\pm$ 0.2 & - & - & - \\
RGG 123$^c$ & S{\'e}rsic & 17.53 $\pm$ 0.11 & 1.00 & 1.41 $\pm$ 0.09 & - \\
\\

& PSF & 19.94 $\pm$ 0.01 & - & - & - \\
RGG 127$^a$ & Inner S{\'e}rsic & 20.40 $\pm$ 0.01 & 0.95 $\pm$ 0.48 & 0.09 $\pm$ 0.02 & 0.53 \\
& Outer S{\'e}rsic & 18.13 $\pm$ 0.05 & 0.70 $\pm$ 0.23 & 1.25 $\pm$ 0.02 & 0.68 & Bar (${\it m_{F110W} = 17.75}$)

\enddata
\tablecomments{Column 1: identification number given in \citet{reines}. Column 2: Components in best-fit GALFIT model. Column 3: Total apparent ST magnitude reported by GALFIT. Column 4: S{\'e}rsic index reported by GALFIT. Column 5: Effective radius reported by GALFIT, converted to kpc. Column 6: Axis ratio (b/a) reported by GALFIT. Column 7: Any additional component included in the best-fit model. 
\tablenotetext{a}{Fitting results from \citet{schutte}. Outer S{\'e}rsic indices were allowed to vary rather than being fixed at $n = 1$.} 
\tablenotetext{b}{Fitting results from \citet{baldassare}. Magnitudes are in the WFC3/IR F160W filter.}
\tablenotetext{c}{Fitting results from \citet{jiang}. Magnitudes are in the WFPC2 F814W filter.}
}
\end{deluxetable*}

\newpage

\bibliography{refs}

\end{document}